\documentclass[11pt, a4paper]{article}
\pdfoutput=1

\usepackage{amsmath,color,multicol}
\usepackage{amsfonts}
\usepackage{amssymb}
\usepackage{graphicx}
\usepackage{mathrsfs}
\usepackage{epsfig}
\usepackage{latexsym}
\usepackage{graphicx}
\usepackage{color}
\usepackage{amsmath,amssymb}
\usepackage{cite}
\usepackage{slashed, cancel}
\usepackage{hyperref}
\usepackage{datetime}
\usepackage{cancel}
\usepackage{multirow}
\usepackage[normalem]{ulem}

\def\circa#1{\,\raise.3ex\hbox{$#1$\kern-.75em\lower1ex\hbox{$\sim$}}\,}

\newcommand{\beq}{\begin{equation}}
\newcommand{\eeq}{\end{equation}}
\newcommand{\beqn}{\begin{eqnarray}}
\newcommand{\eeqn}{\end{eqnarray}}

\def\app#1#2{%
  \mathrel{%
    \setbox0=\hbox{$#1\sim$}%
    \setbox2=\hbox{%
      \rlap{\hbox{$#1\propto$}}%
      \lower1.1\ht0\box0%
    }%
    \raise0.25\ht2\box2%
  }%
}

\numberwithin{equation}{section}

\font\tenrsfs=rsfs10 at 12pt

\font\sevenrsfs=rsfs7
\font\fiversfs=rsfs5
\newfam\rsfsfam
\textfont\rsfsfam=\tenrsfs
\scriptfont\rsfsfam=\sevenrsfs
\scriptscriptfont\rsfsfam=\fiversfs
\def\mathscr#1{{\fam\rsfsfam\relax#1}}

\hypersetup{colorlinks, citecolor=bluscuro, linkcolor=black, urlcolor=bluscuro}
\definecolor{rossos}{rgb}{0.8,0.2,0.3}
\definecolor{bluscuro}{rgb}{0.15, 0.2, .85}
\definecolor{bluchiaro}{cmyk}{1,.3,0.,0.1}

\makeatother   
\newcommand{\DM}{\chi}
\newcommand{\CAP}{\eta}

 \def\be   {\begin{equation}}   \def\ee   {\end{equation}}
 \def\ba   {\begin{array}}      \def\ea   {\end{array}}
 \def\bea  {\begin{eqnarray}}   \def\eea  {\end{eqnarray}}
 \def\bean {\begin{eqnarray*}}  \def\eean {\end{eqnarray*}}

\def\hhref#1{\href{http://arxiv.org/abs/#1}{#1}} 

\begin{document}

%
\vspace{-1cm}

\begin{flushright} 
IPPP/17/8
\end{flushright}

\newcommand{\stau}{\widetilde{\tau}}
\newcommand{\bino}{\widetilde{B}}

\vspace{0.5cm}
\begin{center}

{\LARGE \textbf {Simplified models of dark matter \\ \vspace{3mm} with a long-lived co-annihilation partner
%
}}
\\ [1.cm]
{\large{\textsc{
Valentin V. Khoze$^{\,a\,}$,
Alexis D. Plascencia$^{\,a\,}$
and
Kazuki Sakurai$^{\,a,\,b\,}$
}}}
\\[0.5cm]

\textit{
$^{a}$ Institute for Particle Physics Phenomenology, Department of Physics, \\
Durham University, Durham DH1 3LE, United Kingdom
\\  \vspace{3mm}
$^{b}$ Institute of Theoretical Physics, Faculty of Physics,\\University of Warsaw, ul.~Pasteura 5, PL--02--093 Warsaw, Poland\\[0.1cm]
}
\end{center}

\vspace{0.7cm}

\begin{center}
\textbf{Abstract}
\begin{quote}

We introduce a new set of simplified models to address the effects of 3-point interactions between
the dark matter particle, its dark co-annihilation partner, and the Standard Model degree of freedom, which
we take to be the tau lepton. The contributions from dark matter co-annihilation channels are highly relevant for
a determination of the correct relic abundance. We investigate these effects as well as the discovery potential 
for dark matter co-annihilation partners at the LHC.
A small mass splitting between the dark matter and its partner is preferred by the co-annihilation
mechanism and suggests that the co-annihilation partners may be long-lived (stable or meta-stable) at collider scales. 
It is argued that such long-lived electrically charged particles can be looked for at the LHC 
in searches of anomalous charged tracks. This approach and the underlying models provide an 
alternative/complementarity to the mono-jet and multi-jet based dark matter searches widely used in the context of simplified models with $s$-channel mediators. 
We consider four types of simplified models with different particle spins and coupling structures.
Some of these models are manifestly gauge invariant and renormalizable, others would ultimately require a 
UV completion. These can be realised in terms of supersymmetric models in the neutralino--stau co-annihilation regime,
as well as models with extra dimensions or composite models.
\end{quote}
\end{center}

\def\thefootnote{\arabic{footnote}}
\setcounter{footnote}{0}
\pagestyle{empty}

\newpage
\tableofcontents
\vspace{5mm}

\pagestyle{plain}

\section{Introduction}

\label{sec:introduction}

The existence of dark matter (DM) in the Universe has been established by a number of astrophysical observations based on gravitational interactions. Using the standard model of cosmology, the data collected by the 
Planck mission~\cite{Ade:2015xua} 
implies that DM constitutes nearly 85\% of the total matter content in the universe. Nevertheless, a microscopic
description of DM by a fundamental particle theory is still missing and the nature of dark matter remains largely unknown. 
There is a well-established approach to search for dark matter which relies on the three distinct detection strategies:
the direct detection, the indirect detection and DM searches at colliders.\footnote{ For a classic review 
of particle DM candidates and the experimental search strategies see e.g.~\cite{Bertone:2004pz}.}
The direct detection searches use underground experiments that measure nucleon recoil in order to 
detect the interaction between DM and nucleons. The indirect detection strategy uses experiments that look for an astrophysical 
signal coming from decays or annihilation of DM particles into the Standard Model (SM) particles. Finally, dark matter 
is actively searched at colliders, presently at the LHC, with the aim to produce DM particles in proton collisions.
As the SM does not contain a viable DM candidate, any evidence of DM production at colliders would be a signal
of new physics, the discovery of which is arguably one of the most important goals in the field.

\medskip

Despite an intense experimental effort and surveys of these three directions,
the dark matter has so far proven to be elusive. The no-observation of DM is starting to put some pressure on the 
so-called WIMP Miracle paradigm, which posits that the observed relic abundance can be explained
by DM candidates which are weakly interacting massive particles (WIMPs) with masses in the 10s of GeV to a few TeV range
(assuming simple $2\to 2$  DM annihilation to SM particles and the standard thermal freeze-out mechanism).
A growing number of such WIMP models of DM are being strongly constrained by, or at least show tension with the experimental limits, including supersymmetric DM realisations discussed 
in~\cite{Cheung:2012qy,Baer:2013vpa,Bagnaschi:2015eha,Cohen:2013ama}
as well as other models considered in e.g.~\cite{Cohen:2011ec,Chala:2015ama}.

\medskip  

Our ignorance of the dark sector structure and the negative experimental results for DM searches
have motivated more model-independent studies which fall into two categories. The first is based 
on exploiting effective operators describing the low energy interactions between the DM and the SM 
particles \cite{Beltran:2008xg,
Cao:2009uw,
Goodman:2010yf,
Bai:2010hh,
Fan:2010gt,
Goodman:2010ku,
Zheng:2010js,
Buckley:2011kk,
Yu:2011by,
Rajaraman:2011wf,
Cheung:2012gi,
Cotta:2012nj,
Dreiner:2013vla,
Haisch:2013fla}.
This EFT approach manifestly does not depend on the UV structure of the (unknown) microscopic dark sector theory 
and works well when applied to the low energy experiments, such as the direct detection.
However, the EFT approximation often breaks down when studying collider signatures
since the cut-off of the effective field theory may not be larger than the LHC's energy scale
or the dark sector often requires a new mediator particle other than the DM
which may dramatically alter the collider signature itself
\cite{Busoni:2013lha,
Buchmueller:2013dya,
Busoni:2014sya}.  

\medskip  

The alternative  framework is the simplified model approach, in which
sets of phenomenological models are constructed with a minimal particle content 
to describe various experimental signatures. 
This approach turns out to be very useful and searches for dark matter at colliders are now
commonly described in terms of simplified models with scalar, 
pseudo-scalar, vector and axial-vector 
mediators~\cite{Buckley:2014fba,Harris:2014hga,Haisch:2015ioa,Buchmueller:2014yoa}. 
These simplified models have become the main vehicle for interpreting DM searches at the 
LHC~\cite{Abercrombie:2015wmb,Boveia:2016mrp}
and for projecting the DM reach of future hadron colliders \cite{Golling:2016gvc,Harris:2015kda,Khoze:2015sra}.

\medskip   

These simplified models can be viewed as arising from integrating out the irrelevant particles
and taking a certain limit of the more detailed microscopic theories. The dependence on specific
details of any particular UV embedding in this case is by definition beyond the scope of the simplified models settings.
An interesting question to ask is of course whether and which types of UV completions of specific simplified models 
are possible and if the additional degrees of freedom would affect the simplified model predictions at 
particular collider scales. For recent examples and studies of such `next-to-simplified models' we refer the reader to
Refs.~\cite{Berlin:2015wwa,Baek:2015lna,Kahlhoefer:2015bea,Englert:2016joy,Goncalves:2016iyg,
Baek:2017vzd,Bauer:2017ota}.

\medskip  

The simplified models used by the LHC experiments and aggregated by the ATLAS-CMS DM Forum and the LHC DM Working Group~\cite{Abercrombie:2015wmb,Boveia:2016mrp} 
are conventionally classified based on the type of mediator particles that connect
the DM to the SM particles.
However, this classification may miss an effect of {\it co-annihilation} 
that can be important to determine the DM relic density \cite{Baker:2015qna}.
In the scenario where the co-annihilation is operative, 
a charged (or coloured) particle is introduced in addition to the DM,
which we call the {\it co-annihilation partner}.
Since the interaction between the co-annihilation partner and the SM particles is unsuppressed,  
they annihilate efficiently into the SM particles in the early universe.
Due to the thermal transition between the DM and the co-annihilation partner,
the DM density is also reduced.  
This scenario does not require conventional interactions between the DM and the ordinary particles
through a mediator, and otherwise severe experimental constraints, can easily be avoided. 
Simplified model studies addressing DM co-annihilation
and collider signatures 
so far have mostly focused on the coloured co-annihilation partners \cite{Garny:2012eb,
Kelso:2014qja,
Ibarra:2015nca,
Ellis:2015vaa,
Baker:2015qna,
Buschmann:2016hkc,
Bauer:2016gys}, with only few exceptions as in~\cite{Cirelli:2005uq} (or in~\cite{Khoze:2014xha}
including semi-annihilation effects between two different components of dark matter, e.g.~Vector Vector $\to$ Vector Scalar).

\medskip  

The collider signature is also different in the co-annihilation scenario from the usual DM simplified models.
Since the co-annihilation partner couples to the SM sector with an unsuppressed coupling,
the production rate is much higher for the co-annihilation partners than for DM particles.
Moreover, the co-annihilation partner can be long-lived at colliders because 
its mass difference from the DM mass is small and 
the decay rate incurs a significant phase space suppression.
This may be the case in particular when the co-annihilation partner has a contact interaction
with the DM particle and a $\tau$-lepton,
since if the mass difference is smaller than $m_\tau$,
the co-annihilation partner decays into multi-body final states via an off-shell $\tau$,
leading to a strong phase space suppression. 
This situation is familiar in supersymmetric (SUSY) theories with the stau co-annihilation 
\cite{Ellis:1998kh, 
Jittoh:2005pq, Kaneko:2008re, Jittoh:2010wh,
Citron:2012fg, 
Konishi:2013gda,
Desai:2014uha}.

\bigskip  
\medskip

In this paper, we introduce a class of simplified models that enables us to study the
phenomenology of the dark sector containing a co-annihilation partner.
Inspired in part by the neutralino--stau co-annihilation mechanism in SUSY theories, we want to recreate it in 
more general settings using a new class of simplified model.
In Section {\bf \ref{sec:SMS}} we will define four types of simplified models 
with different particle spins and coupling structures
and assume the existence of  
a contact interaction involving the DM particle, its co-annihilation partner and the SM $\tau$-lepton.
Our simplified model choices include a fermionic DM with a scalar co-annihilation partner, 
a scalar DM with a fermionic co-annihilation partner  
and a vector DM with a fermionic co-annihilation partner.
Some of these models are manifestly gauge invariant and renormalizable, others are supposed
to descend from a more detailed UV complete theory with or without supersymmetry,
some may be realised as a certain limit of composite models,
or descent from models with large extra dimensions.
The expressions for our Simplified Model Lagrangians and the definitions of the free
parameters characterising the models can be found in
Eqs.~\eqref{eq:Lag1}, \eqref{eq:int_2}-\eqref{eq:LagVector} in Section {\bf \ref{sec:res}}.
The Section {\bf \ref{sec:coan}} explains the co-annihilation mechanism for computing the
DM relic density in the context of our simplified models. This is followed by a general overview of 
experimental signatures for direct and indirect detection and collider searches
in Section {\bf \ref{sec:exp}}.
Our main results are presented and discussed in Section {\bf \ref{sec:res}}.
Finally in Section {\bf \ref{sec:conclusions}} we draw our conclusions.


\section{Simplified models}
\label{sec:SMS}

To implement the Dark Matter co-annihilation mechanism we consider dark sectors which include 
two distinct degrees of freedom:
the DM particle, $\DM$, and the charged co-annihilation partner (CAP), $\CAP^{(\pm)}$.
We assume that both of these dark sector particles have odd parity under a $Z_2$ symmetry to ensure
the stability of the dark matter $\DM$. Our simplified models are defined by the three-point interactions
between $\DM$, $\CAP$ and the $\tau$-lepton of the Standard Model sector,
\beqn
{\cal L} \,\supset\, g_{_{\rm DM}} \, \DM \, \CAP \, \tau \,+\, {\rm h.c.}~\,.
\label{eq:op}
\eeqn
Here $g_{_{\rm DM}}$ denotes the dark sector coupling constant which we take to be real and
we also note that $\CAP$ has a non-vanishing $\tau$-lepton number. 
Restricting the particle content of our simplified models to spins not higher than 1,
we consider three possible spin assignments\footnote{
An additional potential assignment ($1 \over 2$, $1$) leads to $\CAP$ being an electrically charged 
vector boson which prevent us from finding an $\rm SU(2)_L \times U(1)_Y$ invariant operator for Eq.~\eqref{eq:op}. 
We therefore will not consider this option further.
}
 for the ($\DM$, $\eta$) pair:
($1 \over 2$, $0$), 
\newline  ($0$, $1 \over 2$) and ($1$, $1 \over 2$).
The corresponding simplified DM-co-annihilation models we wish to consider are
summarised in Table~\ref{tab:models}.
\begin{table}[t!]
  \centering
  \begin{tabular}{|c|c|c|c|c|c|c|}
    \hline  
    \multicolumn{4}{|c|}{\bf Model-1a } \\  
    \hline
    Component & Field & Charge & Interaction \eqref{eq:Lag1} \\
    \hline \hline
     DM & Majorana fermion $(\chi)$ & ~$Y=0$~&\multirow{2}{*}{ ~~$\phi^* (\chi \tau_R) + {\rm h.c.}$~~ } \\
    \cline{1-3}
     CAP& Complex scalar $(\phi)$ & $Y=-1$&\\
    \hline
  \end{tabular}
  \\
\vspace{4mm}
  \begin{tabular}{|c|c|c|c|c|c|c|}
    \hline  
    \multicolumn{4}{|c|}{\bf Model-1b} \\  
    \hline
    Component & Field & Charge & Interaction \eqref{eq:int_2}-\eqref{eq:int_22} \\
    \hline \hline
     DM & Majorana fermion $(\chi)$ & $Q=0$ &\multirow{2}{*}{ $\phi^* (\chi \tau_R)+ \phi^* (\chi \tau_L) + {\rm h.c.}$ } \\
    \cline{1-3}
     CAP& Complex scalar $(\phi)$ & $Q=-1$ &\\
    \hline
  \end{tabular}  
   \\
\vspace{4mm}
  \begin{tabular}{|c|c|c|c|c|c|c|}
    \hline  
    \multicolumn{4}{|c|}{\bf Model-2} \\  
    \hline
    Component & Field & Charge & Interaction \eqref{eq:Lag_2}\\
    \hline \hline
     DM & Real scalar $(S)$ & $Y=0$&\multirow{2}{*}{ ~~$S (\overline{\Psi} P_R \tau) + {\rm h.c.}$~~ } \\
    \cline{1-3}
     CAP& ~~Dirac fermion $(\Psi)$ ~~& $Y=-1$&\\
    \hline
  \end{tabular}    
  \\
\vspace{4mm}
  \begin{tabular}{|c|c|c|c|c|c|c|}
    \hline  
    \multicolumn{4}{|c|}{ \bf Model-3} \\  
    \hline
    Component & Field & Charge & Interaction \eqref{eq:LagVector}\\
    \hline \hline
     DM & Vector $(V_\mu)$ & $Y=0$&\multirow{2}{*}{ $V_\mu (\overline{\Psi} \gamma^\mu P_R \tau) + {\rm h.c.}$ } \\
    \cline{1-3}
     CAP& ~~Dirac fermion $(\Psi)~~$ & $Y=-1$&\\
    \hline
  \end{tabular}    
  \vspace{3mm}
  \caption{Simplified Models of DM with a colourless co-annihilation partner (CAP)}
  \label{tab:models}
\end{table}

\medskip  

{\it A note on notation:}
we use $\chi$ to denote the DM particle and $\eta$ (or $\eta^\pm$) for the co-annihilation particle in general. 
For the simplified models in Table~\ref{tab:models} we have $\chi = \{\chi, \, S, \, V_\mu\}$
and  $\eta = \{\phi, \, \Psi  \}$ depending on the choice of the model.

\medskip

For the ($1 \over 2$, $0$) spin assignment
we consider the case where the dark matter is a Majorana fermion, $\chi$, and the co-annihilation
partner is a complex scalar field, $\phi$,
bearing in mind the similarity of this case with the neutralino--stau co-annihilation picture in SUSY models, where
$\chi$ plays the role of the lightest neutralino, and the scalar $\phi$ is the stau.
In the simplest realisation of this simplified model, which we refer to as the Model-1a in Table~\ref{tab:models}, the Yukawa interactions~\eqref{eq:op}
between the dark sector particles $\chi$,  $\phi$ and the SM involve only the right-handed component
of the $\tau$-lepton, $\tau_R$, hence the co-annihilation scalar $\phi$ is an $\rm SU(2)_L$-singlet. 
At the same time, the second realisation -- the Model-1b -- involves interactions with both left- and right-handed $\tau$-leptons, and hence the stau-like scalar dark partner $\phi$ is charged under the $\rm SU(2)_L$.
The Simplified Model-1a is a UV-consistent theory as it stands; on the other hand, the Model-1b 
should ultimately be embedded into a more fundamental microscopic theory in the UV
to be consistent with the gauge invariance under $\rm SU(2)_L$. One such embedding can for example be a
supersymmetric model with an operational 
neutralino--stau co-annihilation mechanism.

\medskip

The simplified model corresponding to the ($0$, $1 \over 2$) spin assignment
is called Model-2,
in which we introduce a real scalar $S$ as the dark matter
and a Dirac fermion, $\Psi$, as the co-annihilation partner,
assuming they couple together with $\tau_R$.
Model-3 is constructed for the ($1$, $1 \over 2$) spin assignment
that introduces a real vector, $V_\mu$, for the dark matter
and a Dirac fermion, $\Psi$, for the co-annihilation partner,
assuming again the interaction with $\tau_R$.
These two simplified models can be realised in models of extra dimensions
and/or composite models as we will outline in Section~{\bf \ref{sec:res}}. 

\medskip

The simplified models 1a, 2 and 3 constructed above have the following free parameters: 
the dark matter mass, $m_{\rm DM} \,\equiv m_{\DM}$, the mass splitting, $\Delta M \,= M_{\CAP} - m_{\DM}$, and the dark sector coupling, $g_{_{\rm DM}}$.
In Model-1b we fix the dark sector coupling to be the $\rm U(1)_Y$ gauge coupling ($g_{_{\rm DM}} = g^\prime$).
Instead, we introduce the L-R mixing angle, $\theta$, which controls the
relative strength of the coupling to $\tau_L$ and $\tau_R$, as we will discuss later in more detail. 
The simplified model Lagrangians and the parameter definitions are
given in Eq.~\eqref{eq:Lag1} for Model~1a, Eqs.~\eqref{eq:int_2}-\eqref{eq:int_22} for Model~1b, 
Eq.~\eqref{eq:Lag_2} for Model~2 and in~\eqref{eq:LagVector} for Model~3.

%

\section{Co-annihilation}
\label{sec:coan}

\begin{figure}[t!]
\begin{center}
\includegraphics[scale=1.05]{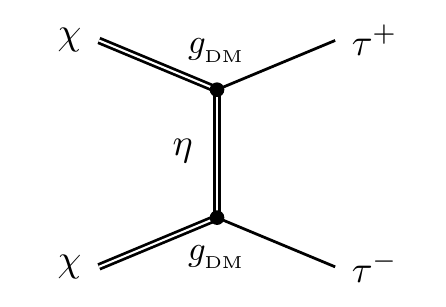} 
\hspace{0mm}
\includegraphics[scale=1.02]{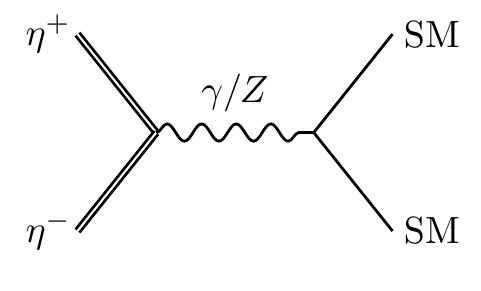} 
\hspace{3mm}
\includegraphics[scale=1.05]{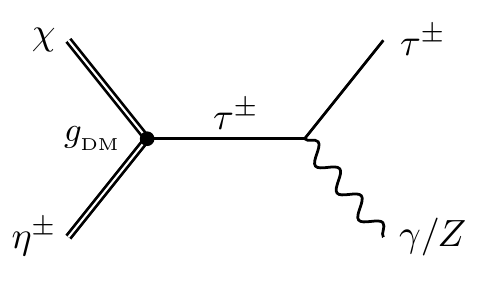} 
\caption{\label{fig:diag} \small
Annihilation and co-annihilation processes.}
\end{center}
\end{figure}

The effect of co-annihilation can be understood qualitatively in the space of simplified model parameters.
First of all, it is worth noting that $\DM$ couples to the SM sector only 
through the operator Eq.~\eqref{eq:op}, whereas $\CAP^\pm$ interacts with the SM particles also via the electromagnetic and weak gauge interactions.
In our simplified models, 
there is a unique channel for the DM pair annihilation:
$\DM \, \DM \to \tau^+ \tau^-$,
as shown in the left diagram in Fig.~\ref{fig:diag}. 
For small $g_{_{\rm DM}}$,
the DM pair annihilation is highly suppressed
because the rate of this process is proportional to $g_{_{\rm DM}}^4$.
For our simplified models 1a,b and 2 where the dark matter is a Majorana fermion or a real scalar ($\chi = \{\chi, \, S\}$),
there is another suppression factor.
The initial state in both these cases forms a spin-0 state (due to the Pauli blocking in the Majorana case).
To conserve the angular momentum, the $\tau^+ \tau^-$ pair in the final state must have 
the opposite chiralities in the $s$-wave contribution, hence 
meaning that this contribution is suppressed by $m_{\tau}^2$ (chiral suppression). 
The dominant contribution then comes from the $p$-wave for a Majorana DM
and $d$-wave for a scalar DM, which are suppressed by the factor $v^2$ and $v^4$, respectively,
where $v$ is the average of the relative velocity of the annihilating DM particles.  
\medskip

Unlike the DM pair annihilation, the annihilation of the CAP particles, 
$\CAP \, \CAP \to SM$ particles, proceeds via the electromagnetic or weak gauge interactions,
as  indicated
in the second diagram of Fig.~\ref{fig:diag}.   As such, the $\CAP \, \CAP$ annihilation can have much 
larger rates than the first process in Fig.~\ref{fig:diag} at small $g_{_{\rm DM}}$.
For a small but non-vanishing values of $g_{_{\rm DM}}$,
there are transition processes between $\eta$ and $\chi$:
$\CAP + {\rm SM} \, \leftrightarrow \, \DM + {\rm SM}$.
These processes are in general much more efficient than annihilation processes,
since the number density of light SM particles is not Boltzmann suppressed
at the time of freeze-out.
As long as the mass splitting, $\Delta M$, is small,
the transition process effectively equalises the number densities of $\DM$
and $\CAP$,
and the DM density (in the unit of the entropy density)
freezes out when the annihilation of $\CAP$ is decoupled.  
We therefore find that in the region of small $g_{_{\rm DM}}$,
the DM relic density is not sensitive to $g_{_{\rm DM}}$
and determined mainly by $\Delta M$ and
$\sigma(\CAP \, \CAP \to {SM\,{\rm particles}}) \times v$.

\medskip  

As $g_{_{\rm DM}}$ approaches the $\rm U(1)_{Y}$ gauge coupling, $g^\prime$,
the co-annihilation process $\DM \, \CAP \to SM\,{\rm particles}$
becomes important (see, for example, the right diagram in Fig.~\ref{fig:diag}).
The rate of this process is proportional to $g_{_{\rm DM}}^2$.
As in the previous process, this process is only effective when $\Delta M$ is small
as we will see below more explicitly.

%
%

\medskip  

For even higher values of $g_{_{\rm DM}}$,
the dark matter pair annihilation, $\DM \, \DM \to \tau^+ \tau^-$,
can become important, since
the annihilation rate is proportional to $g_{_{\rm DM}}^4$.
However, as we have discussed above, for $\chi = \{ \chi, \, S\}$,
this process can never become very large because
it is velocity suppressed.
However it can be dominant for the vector DM case $\chi = V_\mu$.  
Unlike the other channels, the contribution of this process is independent of $\Delta M$.

\medskip  

As it is well known, the DM relic abundance scales as  
\beqn
\Omega_{\rm DM} h^2 
\, \propto \, \langle \sigma_{\rm eff} \,v \rangle^{-1} \,,
\eeqn
where $\langle \sigma_{\rm eff} \, v \rangle$
is the thermal average of the effective annihilation cross-section that is given by \cite{Griest:1990kh}
\beqn
\sigma_{\rm eff} \, v 
&=& \frac{1}{({\tt g}_{\DM} + \overline{{\tt g}}_{\CAP})^2} \,\, \Big[ ~
{\tt g}_{\rm \DM}^2 \cdot \sigma(\DM \, \DM \to \tau^+ \tau^-) +
\nonumber \\ &&~~~~~~~~~~~~~~\,
{\tt g}_{\rm \DM} \overline{{\tt g}}_{\CAP} \cdot  \sigma(\DM \, \CAP \to {SM\,{\rm particles}}) +
\nonumber \\ &&~~~~~~~~~~~~~~~~~
\overline{{\tt g}}_{\CAP}^2 \cdot \sigma(\CAP \, \CAP \to {SM\,{\rm particles}}) 
~ \Big]\,v \,,
\label{eq:sigeff}
\eeqn
with
\beqn
\overline{{\tt g}}_{\CAP} = {\tt g}_{\CAP} \Big( \frac{M_{\CAP}}{m_{\DM}} \Big)^{3/2} \exp\Big(-\frac{\Delta M}{T}\Big) \,,
\eeqn
where ${\tt g}_{\DM}$ and ${\tt g}_{\CAP}$
denote the degrees of freedom of the fields $\DM$ and $\CAP$, respectively,
and should not be confused with the dark sector coupling $g_{_{\rm DM}}$.
Their explicit values are given as 
$({\tt g}_{_S}, \, {\tt g}_{_\chi}, \, {\tt g}_{_\phi}, \, {\tt g}_{_{V_\mu}}, \, {\tt g}_{_\Psi}) \,=\, (1, \,2,\, 2,\, 3,\, 4)$.
Each line of Eq.~\eqref{eq:sigeff} corresponds to
the different contribution discussed above and depicted in Fig.~\ref{fig:diag}. 
The dependence of these contributions on $\Delta M$
can be found through $\overline{{\tt g}}_{\CAP}$. 
Since the freeze-out occurs around $T \sim m_{\rm DM}/25$,
$\Delta M \lesssim m_{\rm DM} /25$
is required in order not to have large suppressions for 
the processes     
$\DM \, \CAP \to {SM\,{\rm particles}}\,$ 
and
$\CAP \, \CAP \to {SM\,{\rm particles}}$.
In this study we are interested in the regime where the co-annihilation is operative,
and we demand $\Delta M$ to be small.
In our numerical study we compute $\Omega_{\rm DM} h^2$  
using \texttt{MicrOMEGAs 4.1.5} \cite{Belanger:2001fz}
implementing the simplified models with help of \texttt{FeynRules 2.0} \cite{Alloul:2013bka} and \texttt{LanHEP 3.2} \cite{Semenov:2014rea}.

\section{Experimental signatures}
\label{sec:exp}

\subsection{Direct detection}

Since the DM couples to the SM sector only through the interaction term Eq.~\eqref{eq:op},
the strength of experimental signatures is rather weak in general for the simplified models
introduced in Section~{\bf \ref{sec:SMS}}.
Direct detection experiments measure the nuclei recoil resulting from their interaction with dark matter,
but such interactions involving DM with quarks and gluons are absent at tree-level in our simplified models.
 At one-loop level, the relevant operators may be generated.
The Higgs mediating contributions are too small because the amplitude is suppressed by the 
product of the tau Yukawa coupling and the Yukawa coupling in the hadron sector.
The relevant operators describing the interactions between the DM and the neutral gauge bosons 
are generated at dimension 6 at the lowest and suppressed by $1/M^2_{\eta}$.
For example, for the Majorana DM case, such an operator is given by
the anapole moment operator ${\cal A} \, \bar \chi \gamma_\mu \gamma_5 \chi \partial^\nu F_{\mu \nu}$.
For $m_{\rm DM} \simeq 500$\,GeV and $\Delta M/m_{\tau} < 1$,
the anapole moment is roughly given by ${\cal A}/g_{_{\rm DM}}^2 \sim 8 \cdot 10^{-7} \,[\mu_N {\cdot \rm fm}]$~\cite{Kopp:2014tsa}, which is more than one order of magnitude smaller than the
current limit obtained by LUX \cite{Akerib:2013tjd} and also smaller than the projected sensitivity of LZ \cite{Malling:2011va}, even for $g_{_{\rm DM}}^2=1$.\footnote{The limits mentioned here assume the observed energy density of the DM.
On the other hand, for $m_{\rm DM} \simeq 500$\,GeV and $g_{_{\rm DM}} \simeq 1$,
all of our simplified models underproduce the $\chi$ particles.
The actual constraints would therefore be even milder if this effect is taken into account.}
Although a dedicated study may shed some light on
the future direct detection prospects for our simplified models,
we shall postpone such a study to a future work.

\subsection{Indirect detection}

Indirect detection experiments are looking for high energy cosmic rays or neutrinos 
originated from the DM pair annihilation (or decay) in the present Universe. 
For the $2 \to 2$ topology, the only relevant process is $\DM \, \DM \to \tau^+ \tau^-$ shown by 
the right diagram of Fig.~\ref{fig:diag}.
As mentioned in the previous section, for $\chi = \{ \chi, \, S \}$ 
this process suffers from the chiral suppression,
and the signal rate for the indirect detection goes below the experimental sensitivity.
The chiral suppression is absent for $\chi = V_{\mu}$ (Model-3).
In Appendix~\ref{ap:2} we compare the annihilation rate of $V_{\mu} \, V_{\mu} \to \tau^+ \tau^-$
with the current limit obtained by Fermi-LAT \cite{Ackermann:2013uma},
taking into account the rescaling of the flux factor by the predicted relic abundance.
We find that the annihilation rate in Model-3 is two orders of magnitude smaller than the current limit
across the parameter region.

\medskip
The $2 \to 3$ scattering, $\DM \, \DM \to \tau^+ \tau^- \gamma$, 
may be more interesting in a small $\Delta M$ region.
In this regime, the reaction rate of this process is enhanced in the following way.
One of the DM particles can be converted into 
a slightly off-shell $\CAP$ radiating off a soft tau,
$\DM \to \CAP^\pm \tau^\mp$.
This $\CAP^\pm$ can then co-annihilate with the other $\DM$ particle
via $\DM \, \CAP^\pm \to \tau^\pm \gamma$ (see, for example, the third diagram in Fig.~\ref{fig:diag}).
Since the converted $\CAP^\pm$ is only slightly off-shell,
the propagator of $\CAP^\pm$ is enhanced, and 
the energy distribution of the produced $\gamma$ has a peak around $m_{\rm DM}/2$,
which can be seen as a bump in a smoothly falling background.
Although this signature is in principle promising,
it has been shown that for $\Delta M \ll m_{\rm DM}$ the annihilation rate is nevertheless
below the experimental sensitivities \cite{Toma:2013bka, Kopp:2014tsa, Giacchino:2013bta, Kumar:2016cum}.
For example, for the Majorana (scalar) DM with $m_{\rm DM} = 500$\,{\rm GeV} and $\Delta M/m_{\tau} < 1$,
the annihilation rate is roughly given by 
$\langle v \sigma(\chi \chi \to \tau^+ \tau^- \gamma)\rangle/g^2_{_{\rm DM}} \sim 5 \cdot 10^{-29} \,(5 \cdot 10^{-28})$ [cm$^3$/s], which is
smaller than the current limits obtained by Fermi-LAT \cite{Ackermann:2013uma} and HESS \cite{Abramowski:2013ax}, and also below the future sensitivity of CTA \cite{Consortium:2010bc,Garny:2013ama} even for $g_{_{\rm DM}} = 1$
and assuming $\Omega_{\chi} h^2 = \Omega_{\rm DM} h^2 \simeq 0.1197$.
As in the direct detection case, we reserve the dedicated study on the prospects of
the indirect detection sensitivity to our simplified models for a future work.

\subsection{Collider searches}

In general, DM particles can be produced in proton-proton collisions at the LHC and 
the experimental collaborations are looking for signatures of such DM production, usually involving mono-
and multi-jets plus missing energy, or alternatively constraining a direct mediator production which could decay back into SM.
In our simplified models of DM with colourless co-annihilation partners, however, no direct DM production processes 
are possible at tree level 
since the DM couples to the SM sector only via the interactions~\eqref{eq:op}. 

\medskip

\begin{figure}[t!]
\begin{center}
\includegraphics[scale=1.2]{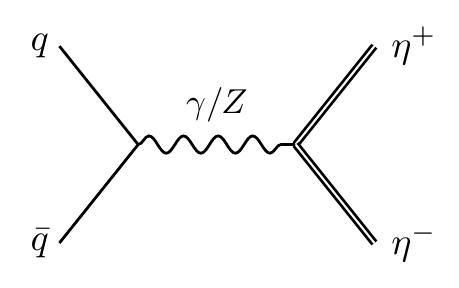} 
\vspace{-3mm}
\caption{\label{fig:prod} \small
Co-annihilation partner (CAP) pair-production process.}
\end{center}
\end{figure}

\begin{figure}[t!]
\begin{center}
\includegraphics[scale=0.55]{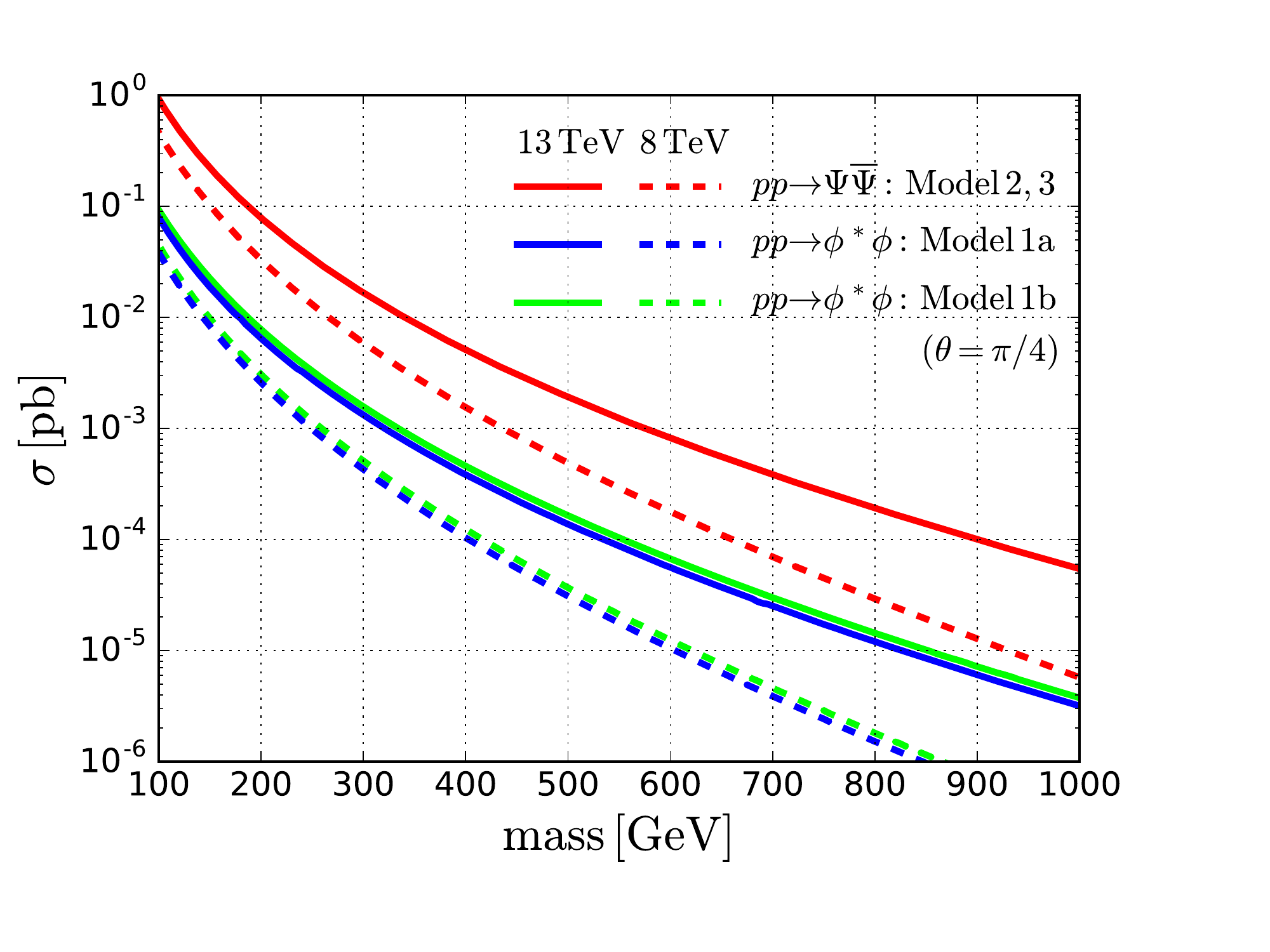} 
\vspace{-3mm}
\caption{\label{fig:xsec} \small
Collider cross-section  $\sigma^{\rm LO} (pp\to \eta^+ \eta^-)$
for the simplified models defined in Table~\ref{tab:models}.}
\end{center}
\end{figure}

Unlike the DM particle, the co-annihilation $\CAP$ particle couples to the SM sector via
electro-weak gauge interactions,
and $\CAP$ can be pair-produced by
exchanging off-shell neutral gauge bosons 
 $q \bar q \to (\gamma/Z)^* \to \CAP \CAP$
as depicted in Fig.~\ref{fig:prod}.
%
%
%
%
%
%
%
The production rate is independent of $g_{_{\rm DM}}$ and is well-defined 
once the mass and quantum numbers of $\CAP$ are specified.
For our simplified models of DM with co-annihilation partners $\CAP$, the latter are either 
a complex scalar or Dirac fermions. 
The $\CAP$ production cross-sections $pp \to \CAP \CAP$ at the 8 TeV and 13 TeV LHC computed at leading order 
by {\tt MadGraph~5} \cite{Alwall:2014hca}
for our range of simplified models are plotted in Fig.~\ref{fig:xsec} as the function of the co-annihilation partner mass. 
It can be seen that the production cross-section in the fermion case is one order of magnitude higher than in the scalar case. 
This is because the scalar production suffers from velocity suppression near the threshold; we will further comment on this effect in Section {\bf 5.3}.
%
%
%
%

\medskip

In the region where the co-annihilation is operative,
$\Delta M$ is small and the decay products of $\CAP$ 
will be too soft to be reconstructed.\footnote{ The LHC phenomenology
of a similar model in the opposite limit ($\Delta M \sim m_{\rm DM}$) have been studied in \cite{Yu:2014mfa}. 
}
The standard strategy to trigger such events is 
to demand additional hard jet originated from the initial state QCD radiation.
This leads to a distinct mono-jet plus large missing energy signature
and the signal can (in favourable settings) be separated from the background.
It is known that the mono-jet channel is powerful 
if $\CAP$ has a colour charge, but for   
our colour-neutral $\CAP$ this prospect is, as one would expect, quite pessimistic.
For example, the study presented in \cite{Low:2014cba}
did not find any limit on the stau mass 
in the stau co-annihilation region in SUSY models 
using a mono-jet channel even for a 100 TeV $pp$ collider with a 3 ab$^{-1}$ integrated luminosity.
In this paper we focus on the sensitivity at the LHC 
and aim to look for an alternative search channel.  

\medskip

%
\begin{figure}[t!]
\begin{center}
\includegraphics[scale=1.]{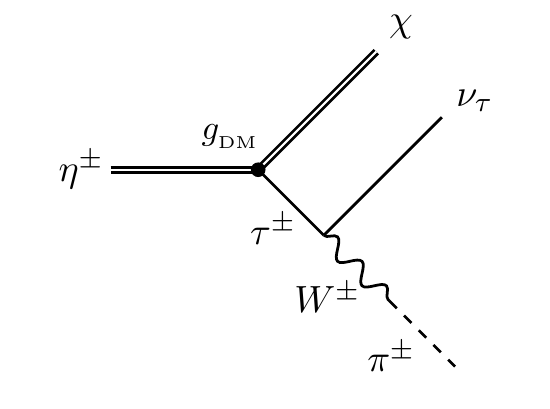} 
\hspace{3mm}
\includegraphics[scale=1.]{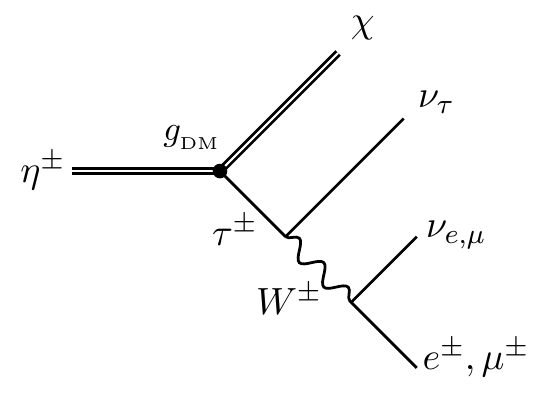} 
\vspace{-2mm}
\caption{\label{fig:dec_diag} \small
The 3-body and 4-body  $\CAP$-decays via an off-shell $\tau$ (and $W)$.}
\end{center}
\end{figure}
%
%
%
\begin{figure}[t!]
\begin{center}
\includegraphics[scale=0.5]{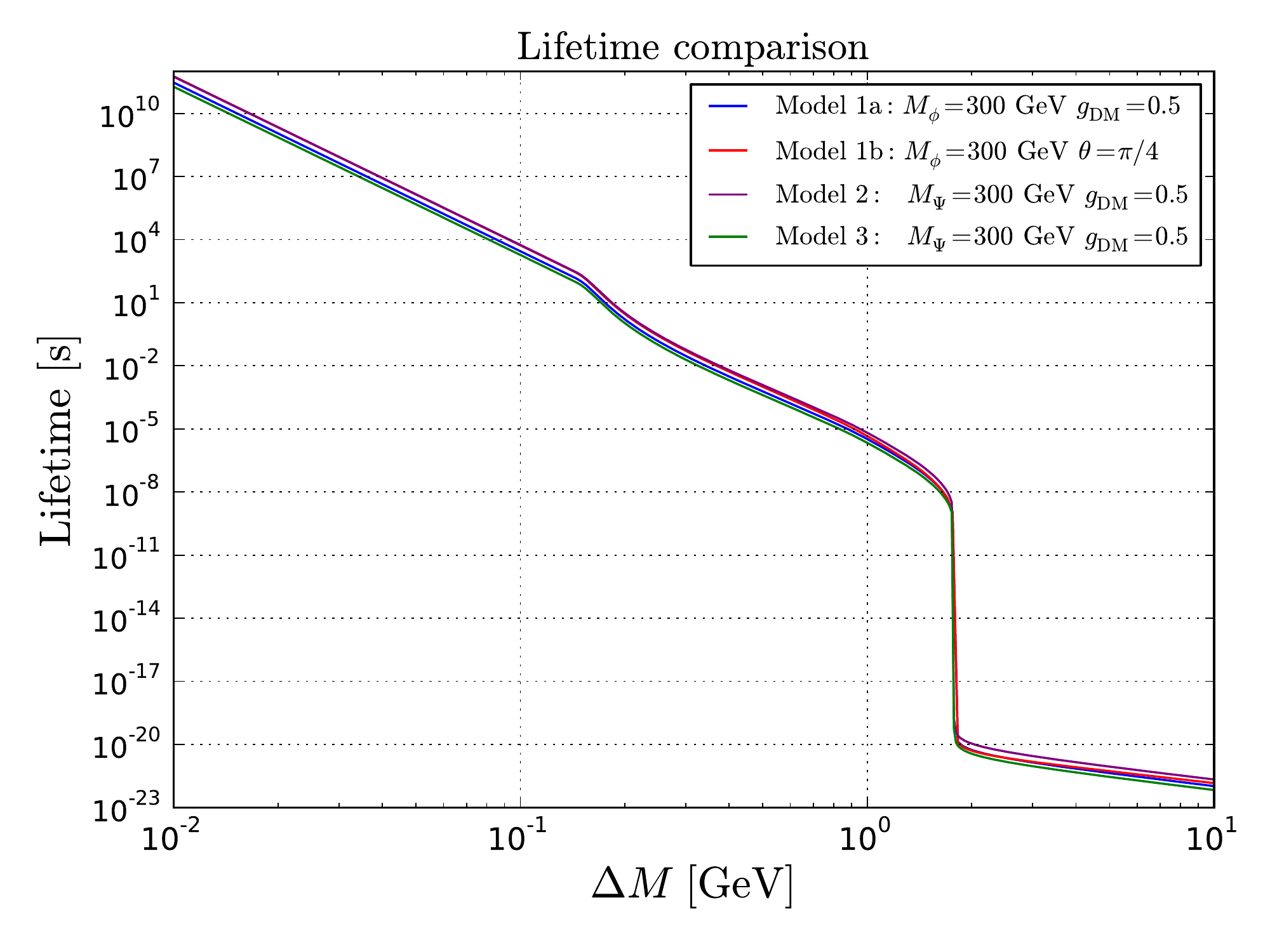} 
\vspace{-2mm}
\caption{\label{fig:lifetime} \small
The lifetime of the co-annihilation partner $\eta^\pm$ as a function of 
the mass splitting $\Delta M = M_\eta - m_\chi$. 
Model~1a (blue): $M_\phi=300{\rm \,GeV}$, $g_{\rm DM}=0.5$,
Model~1b (red): $M_\phi=300{\rm \, GeV}$, $\theta=\pi/4$,
Model~2 (purple): $M_\Psi=300{\rm \, GeV}$, $g_{\rm DM}=0.5$,
Model~3 (green): $M_\Psi=300{\rm \, GeV}$, $g_{\rm DM}=0.5$.  
}
\end{center}
\end{figure}
%

As we have seen in Section~{\bf \ref{sec:coan}},
the effective co-annihilation mechanism in the dark sector imposes an upper bound on the mass splitting
between the DM and the CAP particles, $\Delta M \lesssim m_{\rm DM}/25$.
Furthermore, if $\Delta M$ becomes smaller than the $\tau$-lepton mass, $m_{\tau} = 1.777$ GeV,
the on-shell 2-body decay, $\CAP^\pm \to \DM \tau^{\pm}$,
is kinematically forbidden
and the 3- and 4-body decay modes, $\CAP^\pm \to \DM \, \nu_\tau \, \pi^{\pm}$
and $\CAP^\pm \to \DM \, \nu_\tau \, \ell^\pm \, \nu_\ell$ ($\ell = e, \mu$)
shown in Fig.~\ref{fig:dec_diag},
become dominant.
Since these 3- and 4-body decays are suppressed by the off-shell intermediate 
propagators and the multi-body phase space,
the $\CAP$ decay rate becomes minuscule.

\medskip

We show in Fig.~\ref{fig:lifetime} 
the lifetimes of $\CAP^\pm$ computed with {\tt CalcHEP} \cite{Belyaev:2012qa}
as functions of $\Delta M$
for our simplified models of DM with a co-annihilation partner.
As can be seen, the lifetimes quickly increase
once $\Delta M$ crosses $m_{\tau}$ from above
and reach $\sim 1 \mu s$ around $\Delta M \sim 1$ GeV,
for all simplified models.
If the lifetime is of the order of $\mu s$,
$\CAP$ can reach the tracker and may leave 
anomalously highly ionizing tracks
or slowly moving charged particle signature.
Such exotic charged track signatures are intensively looked for 
by ATLAS \cite{Aaboud:2016dgf, Aaboud:2016uth} and CMS \cite{Khachatryan:2016sfv, Khachatryan:2015lla}
and also can be investigated by the MoEDAL experiment~\cite{Acharya:2014nyr}. 
We calculate the projected limits obtained from anomalous charged track searches
for various simplified models 
and discuss an interplay with the dark matter relic abundance 
obtained by the co-annihilation mechanism in the next section.

\section{Results}
\label{sec:res}

\subsection{Model 1a: Majorana fermion dark matter}

The first simplified model we consider has a Majorana fermion singlet dark matter, 
$\DM = \chi$, and a complex scalar co-annihilation partner,
$(\CAP^+,\CAP^-)  = (\phi^*, \phi) = (\phi^+, \phi^-)$. 
We extend the SM Lagrangian as:
\beqn
\mathcal{L} & = & \mathcal{L}_{\rm SM} \,+\, \mathcal{L_{\rm DM}} \,+\, \mathcal{L_{\rm CAP}} \,+\, \mathcal{L_{\rm int}} \,,
\nonumber \\
\mathcal{L_{\rm DM}} & = & 
\frac{1}{2} \chi (i\slashed{\partial} - m_{\rm DM}) \chi \,,
\nonumber \\
\mathcal{L_{\rm CAP}} & = & 
|D_\mu \phi|^2 - M_\phi^2 \, |\phi|^2 \,,
\nonumber \\
\mathcal{L_{\rm int}} & = & 
g_{_{\rm DM}} \, \phi^* \chi \tau_R \,+\, {\rm h.c.} \,,
\label{eq:Lag1}
\eeqn
where $M_\phi = m_{\rm DM} + \Delta M$
and the covariant derivative $D_\mu$ contains the ${\rm U(1)_Y}$ gauge field. 
This simplified model has a particular interest since 
it can be realised 
in SUSY models by identifying $\chi$ as the Bino
and $\phi$ as the right-handed stau.
We, however, stress that
the model is also interesting on its own right because it is gauge invariant and renormalizable.  
The searches at LEP have already excluded charged particles with mass 
below~$\simeq 100$~GeV~\cite{Heister:2002jca, Achard:2003ge, Abdallah:2003xe}, and we focus on the region with 
$M_{\phi} \gtrsim 100$ GeV. 

\medskip

\begin{figure}[t!]
\begin{center}
\hspace{-5mm}
\includegraphics[scale=0.43]{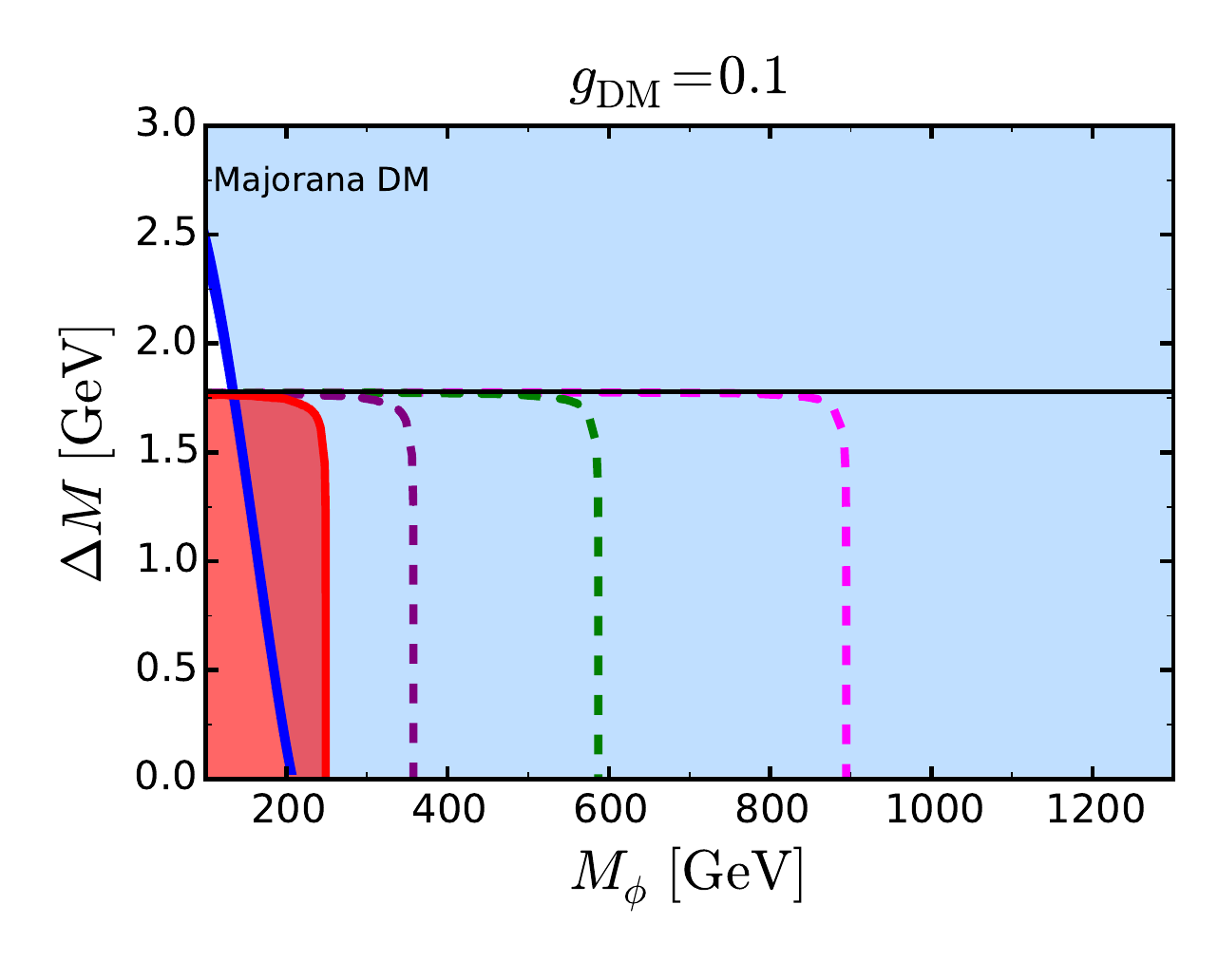}
\includegraphics[scale=0.43]{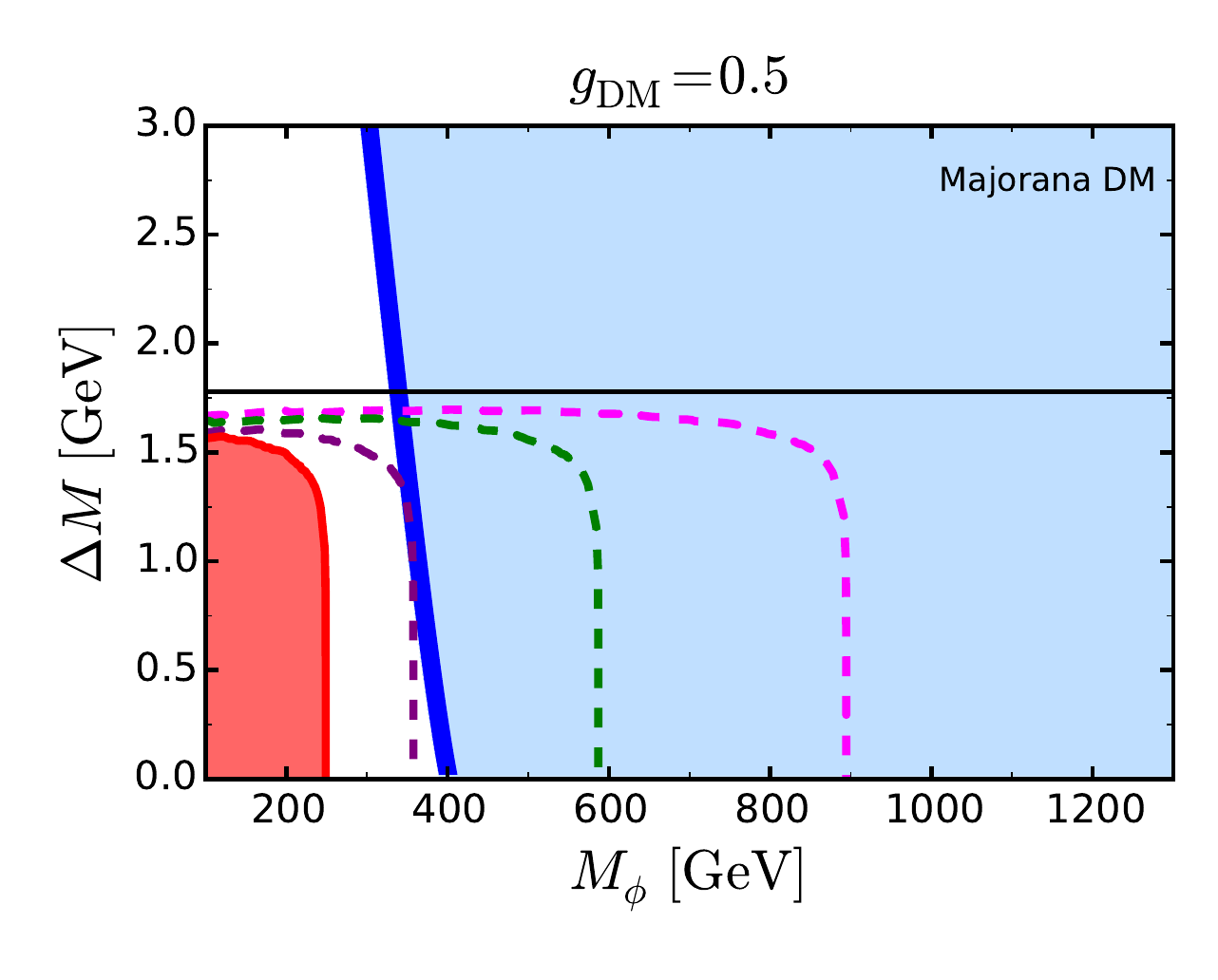}
\includegraphics[scale=0.43]{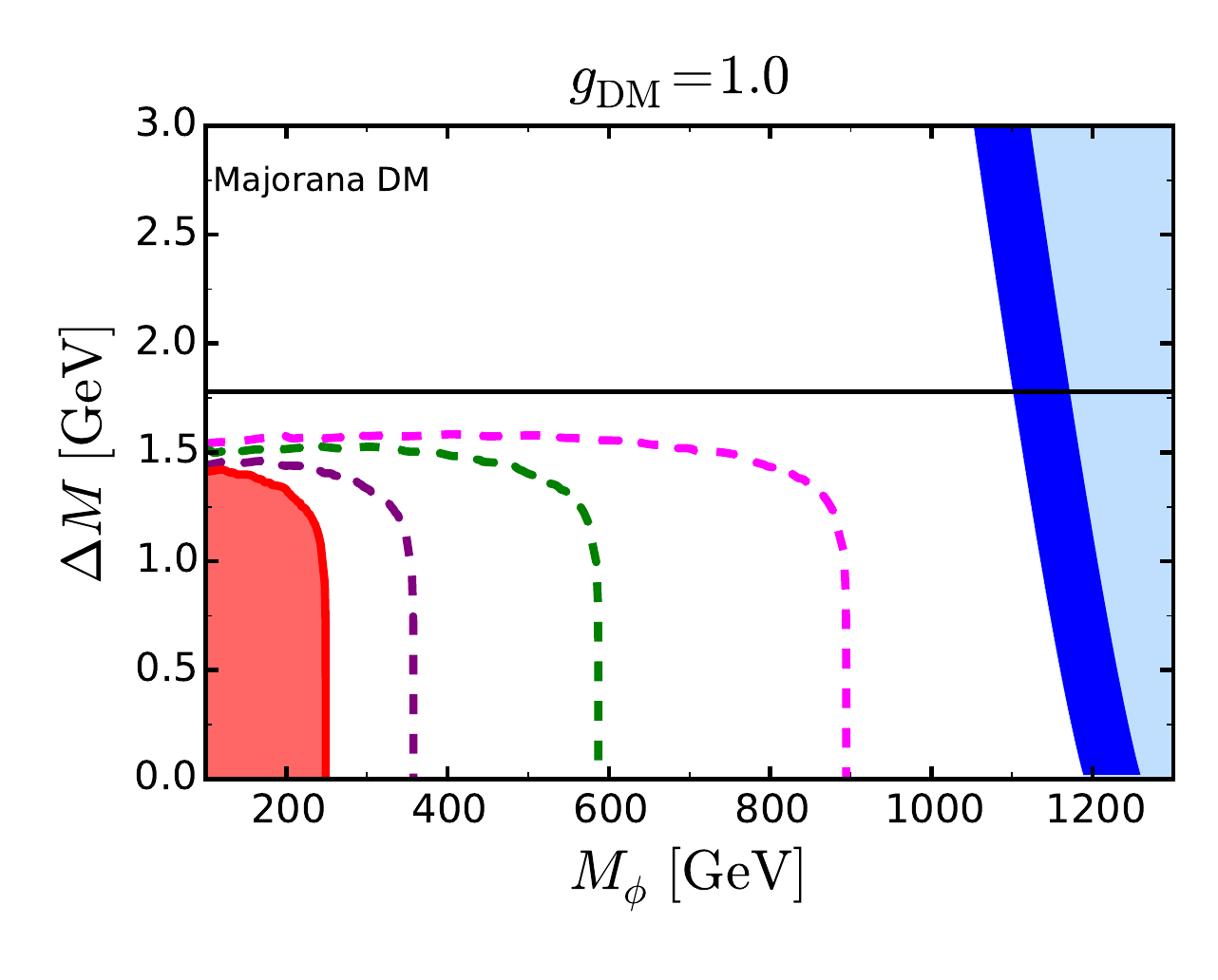}
\vspace{-8mm}
\caption{\label{fig:MajoranaDM} 
\small The co-annihilation strip and collider searches for Majorana DM and a long-lived charged scalar in the Simplified Model~1a. The dark-blue region satisfies the correct dark matter relic abundance within $3\sigma$, the light-blue region overproduces the dark matter energy density. The horizontal black line indicates the mass of the $\tau$ lepton. The region coloured in red corresponds to current HSCP limits at the LHC for center-of-mass energy of 8 TeV and 18.8 fb$^{-1}$. The three dashed lines (purple, green and magenta) correspond to our projections for center-of-mass energy of 13 TeV and 30, 300 and 3000 fb$^{-1}$ of integrated luminosity respectively.}
\end{center}
\end{figure}

We show our numerical results for the Simplified Model~1a in Fig.~\ref{fig:MajoranaDM}. The three plots 
correspond to
different values of the dark matter coupling: $g_{_{\rm DM}} = 0.1$, 0.5 and 1.0
from left to right. The dark-blue region satisfies the correct dark matter relic abundance within $3\sigma$, 
and the light-blue area to the right of it gives a relic abundance which exceeds the observed value and overcloses the universe.
The red region corresponds to
the current 95\% CL excluded region obtained by
the heavy stable charged particle (HSCP) searches
at the LHC
using 8 TeV data with 18.8 fb$^{-1}$ integrated luminosity \cite{Khachatryan:2015lla}.
The contours bounded by the purple, green and magenta dashed lines (from left to right) 
are projected limits 
assuming 13 TeV LHC with 30, 300 and 3000 fb$^{-1}$ integrated luminosities,
respectively.
These projections are obtained by starting with the analysis conducted by CMS \cite{Khachatryan:2015lla}
of the 8 TeV data, and interpolating it to higher energies and luminosities following the Collider Reach method \cite{colliderreach}.\footnote{
    A fast recasting method for a HSCP search has been proposed in \cite{Heisig:2015yla}.  
   We opt for the Collider Reach method, since our main focus is
   to extrapolate the existing limit to higher energies and luminosities.
}
We validated our computational approach by  reproducing the 8 TeV limit on the long-lived stau calculated in~\cite{Evans:2016zau}.
The limit can also be presented as a function of the lifetime and mass of $\phi$.
Such limits are given in Appendix \ref{ap:1}.

\medskip

In Fig.~\ref{fig:MajoranaDM}, the horizontal line represents $\Delta M = m_\tau$.
One can see, as expected, that 
the limit from the HSCP searches is absent if $\Delta M > m_\tau$
since $\phi^\pm$ decays before reaching the tracker.
Once $\Delta M$ gets smaller than $m_\tau$, the propagation path of the $\phi$ charged scalar
$c \tau_\phi$ reaches and then exceeds the detector scale, ${\cal O}(100)$ cm,
although the exact $\Delta M$ needed for exclusion
depends also on $g_{_{\rm DM}}$ since the lifetime
is inversely proportional to $g_{_{\rm DM}}^2$.  
For $g_{_{\rm DM}} = 0.1$,
the HSCP searches can have strong sensitivities as far as 
$\Delta M < m_{\tau}$,
whilst 
$\Delta M \lesssim 1.5$ GeV is required 
for $g_{_{\rm DM}} = 0.5$ and 1.
The model can be constrained at the LHC only when there is a large production cross-section
for $pp \to \phi^+ \phi^-$.
The sensitivity of the HSCP search therefore has a strong dependence on $M_\phi$.
If $\Delta M < 1.3$ GeV, $M_\phi < 240$ GeV is already ruled out by the current data,
and the 95\% CL projected limits are estimated as $M_\phi < 330$, 580 and 870 GeV
for 13 TeV LHC with 30, 300 and 3000 fb$^{-1}$ integrated luminosities, respectively.
These limits are almost independent of $g_{_{\rm DM}}$ and $\Delta M$ as long as $\Delta M < 1.3$ GeV.

\medskip

We have also shown the constraints from the DM relic density in the same plots. 
The dark-blue strip in Fig.~\ref{fig:MajoranaDM}
represents the region where 
the DM relic density,
computed by \texttt{MicrOMEGAs 4.1.5} \cite{Belanger:2001fz},
is consistent with the 
latest Planck satellite measurement 
$\Omega_{\rm DM} h^2 = 0.1197 \pm 0.0022$ \cite{Ade:2015xua}
within the 3-$\sigma$ level.
Note that the DM is overproduced 
on the right of the dark-blue strip, where this region is shaded with light-blue.
Conversely, the DM is underproduced on the left of the dark-blue strip.
This region may not be excluded phenomenologically 
since there may be another component for the DM, whose relic density makes up the remaining part of the $\Omega_{\rm DM} h^2$.
We can therefore identify the white region as the currently allowed region
by the LHC and the DM relic density constraints. 

\medskip

\begin{figure}[t!]
\begin{center}
\includegraphics[scale=0.57]{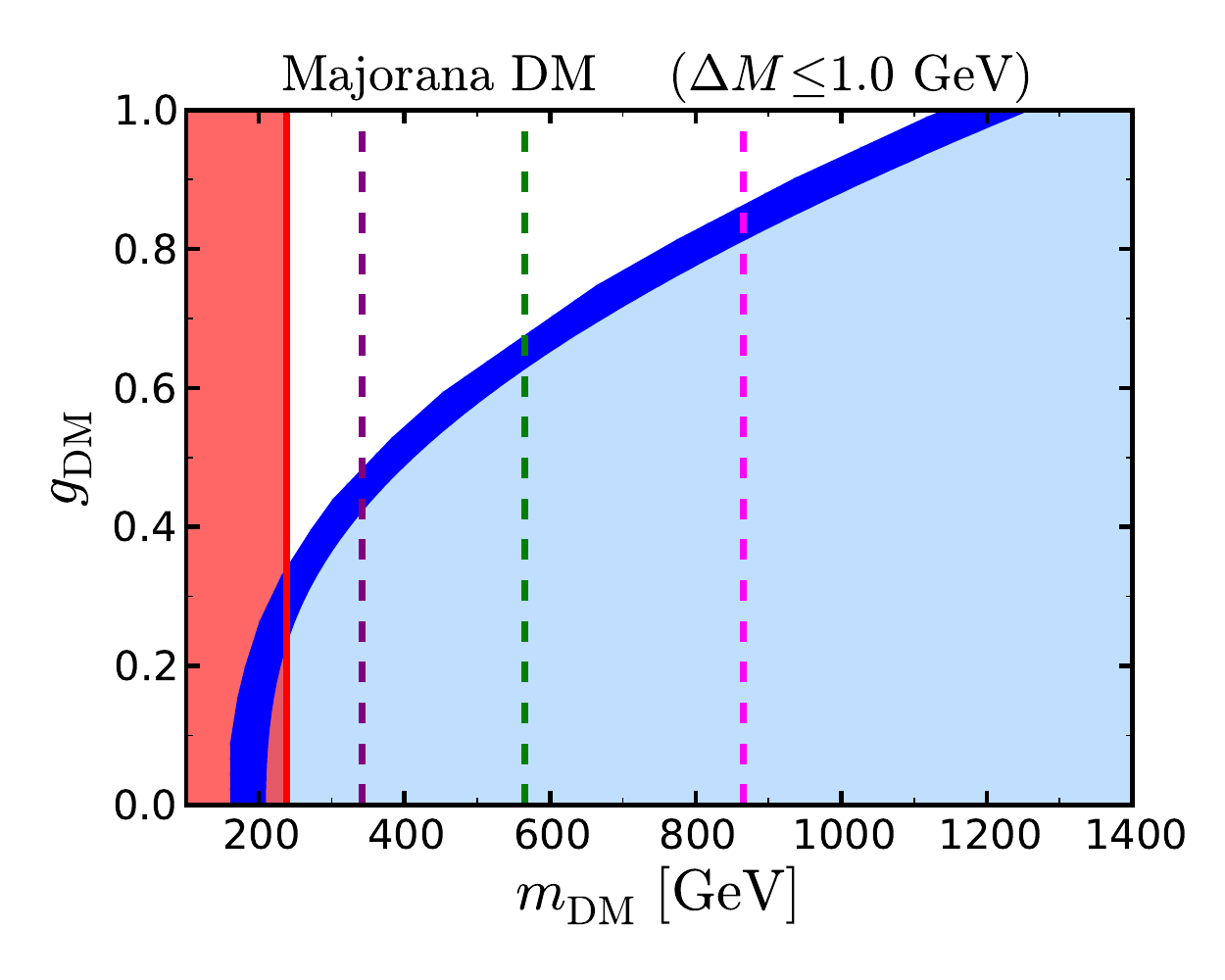}
\vspace{-3mm}
\caption{\label{fig:gRmDM} \small Model~1a:
Plot of the coupling $g_{_{\rm DM}}$versus the dark matter mass $m_{\rm DM}\!=\!m_{\chi}$. We scan over $\Delta M \le 1$~GeV, where $\Delta M\!=\!M_{\phi}\!-\!m_{\chi}$, this is the mass region where the HSCP limits are independent of the coupling $g_{_{\rm DM}}$. The dark blue band satisfies the correct DM relic abundance within $3\sigma$, the region in light blue overproduces the amount of DM. The colour-coding for the exclusion regions is the same as in the previous Figure.}
\end{center}
\end{figure}

As we have discussed in Section {\bf \ref{sec:coan}},
the relic density depends on $\Delta M$ through the co-annihilation mechanism,
which can be seen clearly in Fig.~\ref{fig:MajoranaDM}.
The mass and the dark sector coupling also affect the value of the relic density.
To investigate this behaviour in more detail, in Fig.~\ref{fig:gRmDM} we present a scan of the
 ($g_{_{\rm DM}}$, $m_{\rm DM}$) plane in our Simplified Model~1a over the mass splittings in the region
 $0\le \Delta M \le 1$~GeV.
The dark-blue strip gives the correct relic density within 3$\sigma$.
As previously discussed,
the dependence on $g_{_{\rm DM}}$ is weak if $g_{_{\rm DM}} \ll 1$,
since the $\langle \sigma_{\rm eff} v \rangle$ is almost entirely determined by 
the $\phi^+ \, \phi^- \to {SM\,{\rm particles}}$,
which is independent of $g_{_{\rm DM}}$.
Once $g_{_{\rm DM}}$ gets as large as the $U(1)_{\rm Y}$ gauge coupling,
the second process, $\phi^\pm \chi \to {SM\, {\rm particles}}$,
becomes important, and the dependence on $g_{_{\rm DM}}$ enters into $\Omega_{\rm DM} h^2$.
For very large $g_{_{\rm DM}}$,
the process $\phi^+ \phi^+ \to \tau^+ \tau^+$ (and its conjugate), exchanging $\chi$ in the $t$-channel, becomes dominant 
since it does not incur the chiral suppression and the cross-section is proportional to $g_{_{\rm DM}}^4$.
Because the DM relic density is inversely proportional to $\langle \sigma_{\rm eff} v \rangle$,
the constraint of the DM overproduction excludes small $g_{_{\rm DM}}$ regions
depending on $m_{\rm DM}$.
From this plot we conclude that the high luminosity LHC at 3000 fb$^{-1}$ 
can explore almost the entire region with $g_{_{\rm DM}} \lesssim 1$
except for a small segment around $g_{_{\rm DM}} \sim 0.9$, $m_{\rm DM} \sim 1$ TeV.

\subsection{Model 1b: Effect of L-R mixing}
\label{sec:susy}

In SUSY models we often encounter the situation where
the DM and the lighter stau, $\widetilde \tau_1$ (co-annihilation partner), interact with both left and right-handed $\tau$-leptons
via the L-R mixing in the stau sector. 
To study this case, we extend the previous simplified model such that the co-annihilation partner $\phi$ can couple to both $\tau_L$ and $\tau_R$. 
We will now construct our simplified model 
by starting initially with the $\rm SU(2)_L \times U(1)_Y$ invariant formulation
involving a minimal particle content required for the DM fermion, the co-annihilation scalar(s), and the SM leptons.
We thus introduce a scalar SU(2) doublet $\Phi_L^T = (\phi_\nu, \phi_L)$ and a singlet $\phi_R$ 
with the same hyper-charges as those of the SM doublet $l_3^T = (\nu_\tau, \tau_L)$ and the singlet $\tau_R$, respectively.
We then write down their Yukawa interactions with the DM Majorana fermion $\chi$ as follows,
\beqn
\sqrt{2} \,  g^\prime \, Y_l \, \Phi_L^\dagger \, \chi \, l_3 \,+\, \sqrt{2} \, g^\prime \, Y_e \, \phi_R^* \, \chi \, \tau_R \,+\, {\rm h.c.}\,,
\label{eq:int_1}
\eeqn
where $g^\prime \simeq 0.36$ is the $\rm U(1)_Y$ gauge coupling and $Y_l = -{1 \over 2}$ and $Y_e = 1$ are the corresponding hyper-charges.
These terms are analogous to the bino--stau--tau interaction in SUSY models.  

\medskip

After the electroweak symmetry breaking, the scalars  $\phi_L$ and $\phi_R$ will generically mix with each other 
forming two mass eigenstates, the lighter of which,
\beqn
\phi \,=\, \cos \theta \,\, \phi_L \,+\, \sin \theta \,\, \phi_R\,,
\eeqn
we identify as the co-annihilation particle of our simplified model. The mixing angle $\theta$ will be a free parameter 
in the simplified model. 
After integrating out the heavier scalar eigenstate, the interaction terms in Eq.~\eqref{eq:int_1} reduce to the 
simplified model interaction
\beq
{\cal L}_{\rm int} \,=\, g_L \, \phi^* \chi \tau_L \,+\, g_R \, \phi^* \chi \tau_R \,+\, {\rm h.c.} \,, \label{eq:int_2}
\eeq
with the two couplings given by
\beq
g_L = \frac{1}{\sqrt{2}} g^\prime  \cos \theta, \hspace{4mm}    g_R = - \sqrt{2} g^\prime \sin \theta \,.  \label{eq:int_22}
\vspace{3mm}
\eeq
In the same way, the interaction of $\phi$ with $\gamma$, $Z$ and $W^\pm$ can be obtained  
by extracting $\phi$ from the kinetic terms $|D_\mu \Phi_L|^2 + |D_\mu \phi_R|^2$. 
This defines our Simplified Model~1b, which is determined in terms of three free parameters: $\theta$, $M_\phi$ and $\Delta M= M_\phi-m_\chi$.

\medskip

\begin{figure}[t!]
\begin{center}
\hspace{-5mm}
\includegraphics[scale=0.43]{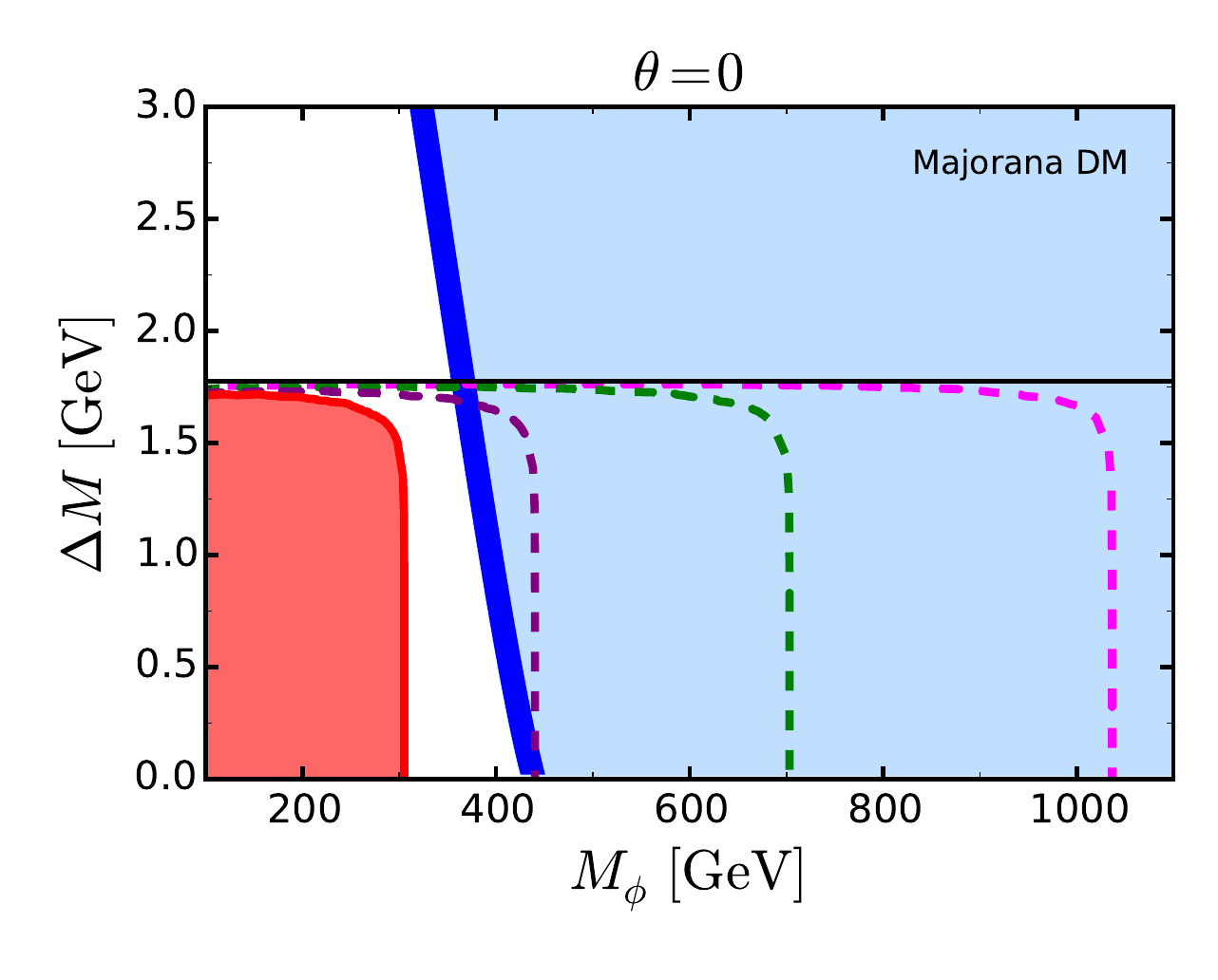}
\includegraphics[scale=0.43]{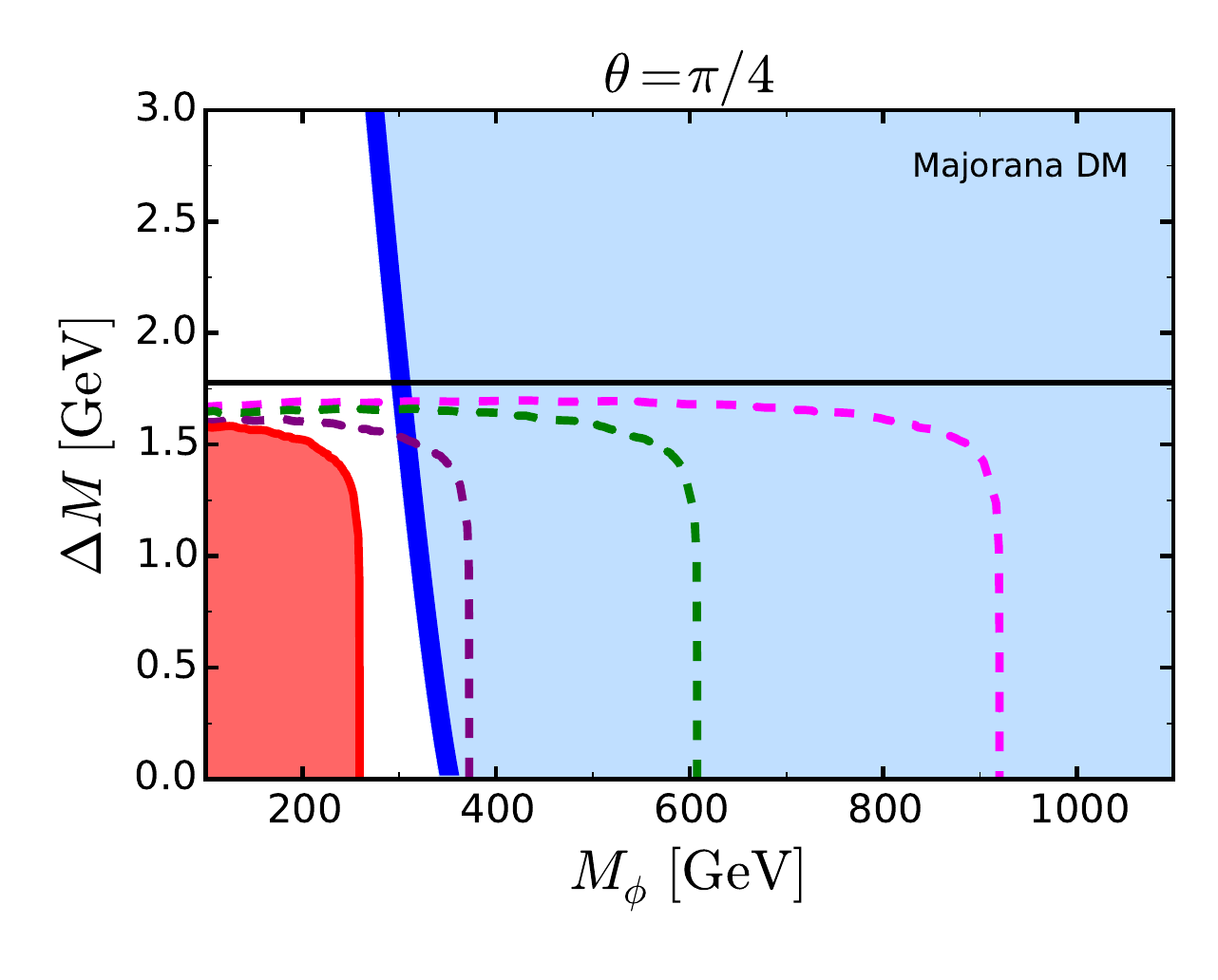}
\includegraphics[scale=0.43]{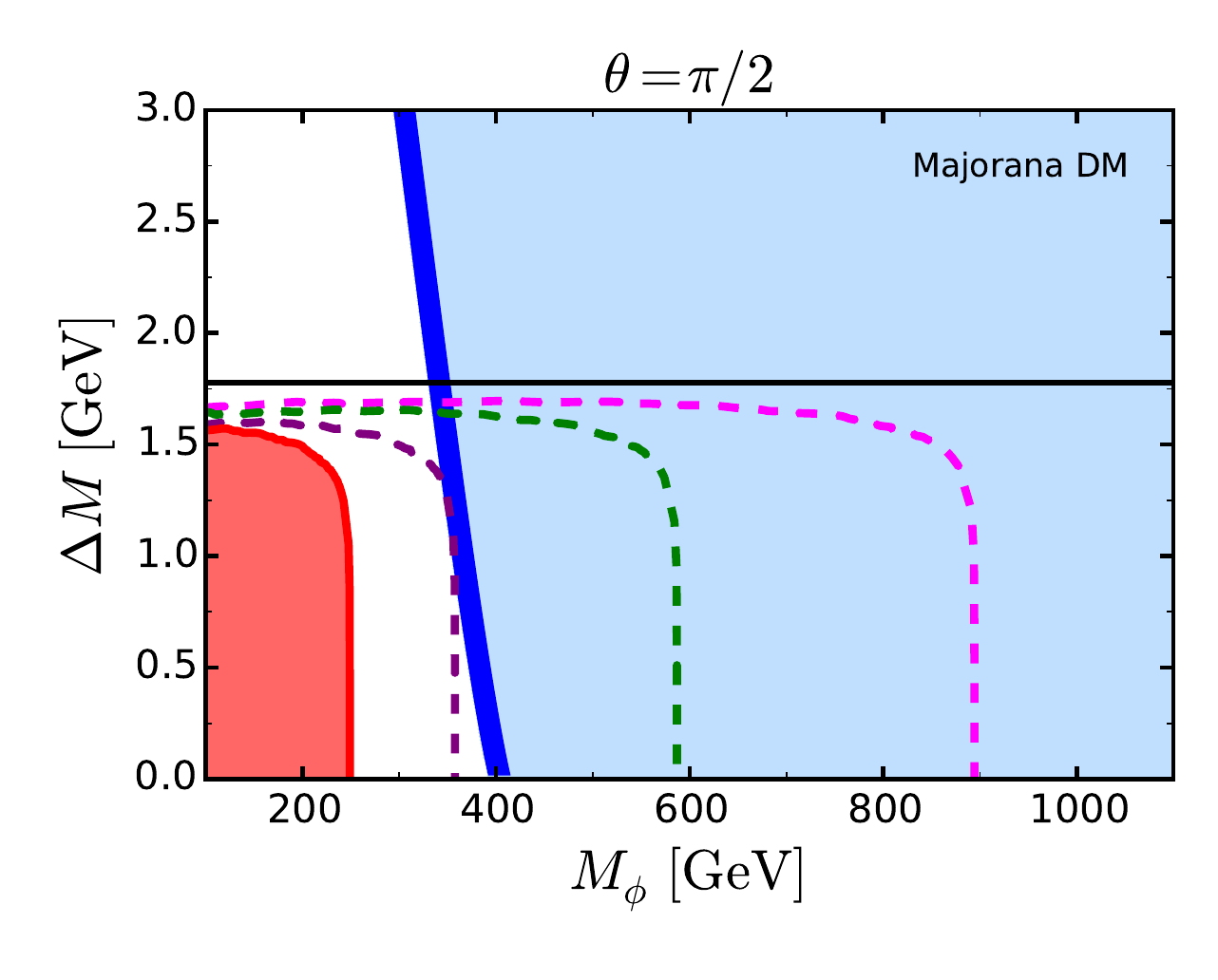} 
\vspace{-3mm}
\caption{\label{fig:susyLimits} \small Model~1b:
$\phi-\chi$ co-annihilation strip and collider searches. The dark-blue region satisfies the correct dark matter relic abundance within $3\sigma$, the light-blue region overproduces the dark matter energy density. The horizontal black line corresponds to the mass of the $\tau$ lepton. The region coloured in red corresponds to current HSCP limits for center-of-mass energy of 8 TeV and 18.8 fb$^{-1}$. The three dashed lines (purple, green and magenta) correspond to our projections for center-of-mass energy of 13 TeV and 30, 300 and 3000 fb$^{-1}$ of integrated luminosity respectively. }
\end{center}
\end{figure}

\begin{figure}[t!]
\begin{center}
\includegraphics[scale=0.45]{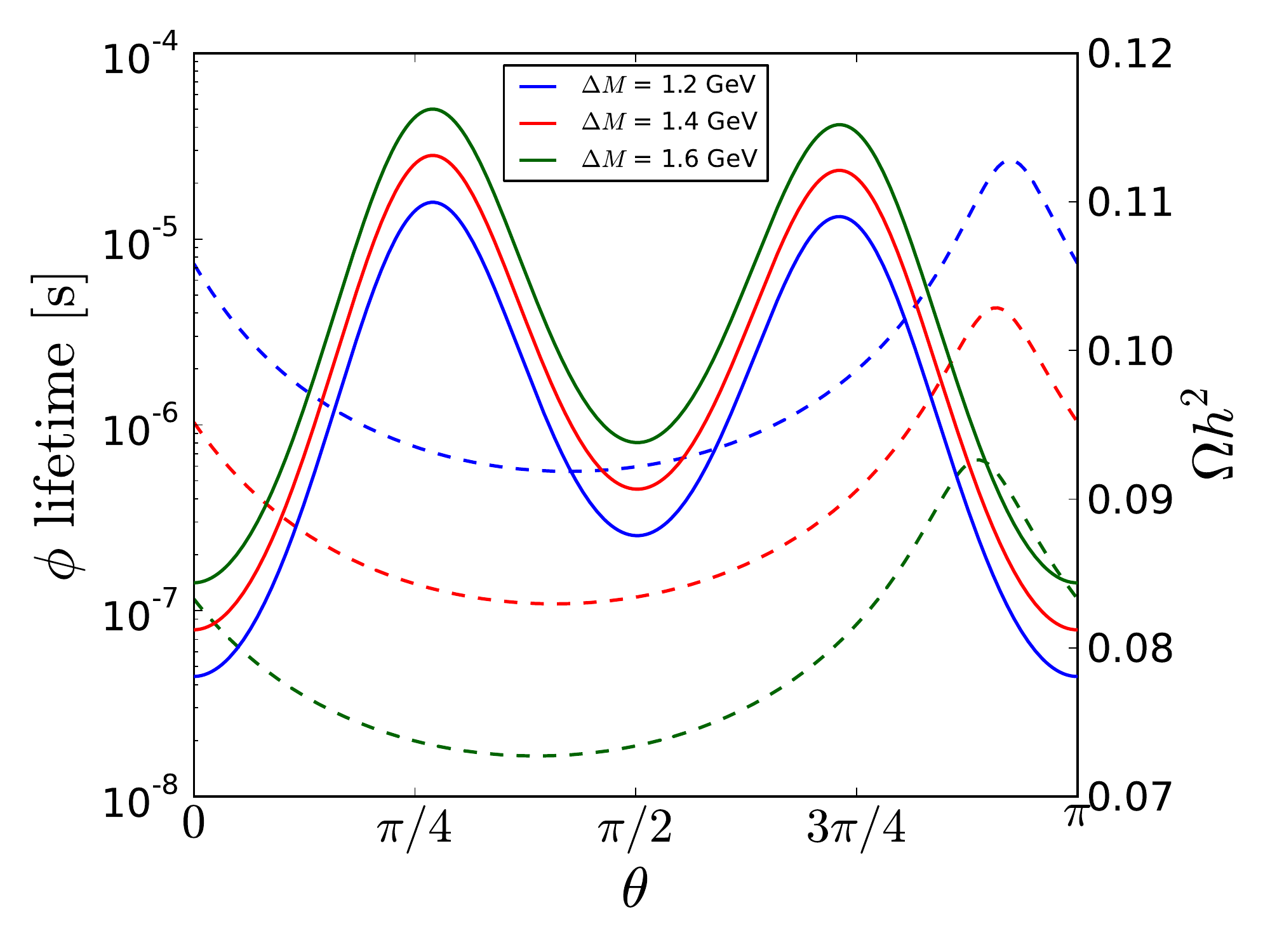}
\vspace{-3mm}
\caption{\label{fig:theta} 
\small
The lifetime of $\phi^\pm$ (dashed) and the DM relic density $\Omega h^2$ (solid)
as functions of the L-R mixing parameter $\theta$.
The DM mass is fixed at 300 GeV and $\Delta M$ is varied as
1.2 (blue), 1.4 (red) and 1.6 (green) GeV.
}
\end{center}
\end{figure}

We show in Fig.~\ref{fig:susyLimits} the constraints in the ($M_\phi$, $\Delta M$) plane for the Simplified Model~1b
for the following parameter choices: $\theta = 0$ for $\phi = \phi_L$ (left plot), $\theta =\pi/4$ for $\phi = (\phi_L + \phi_R)/\sqrt{2}$
(central plot) and $\theta =\pi/2$ for $\phi = \phi_R$ (plot on the right).
We note that $\theta = \pi/2$ corresponds to Model-1a with 
$|g_{_{\rm DM}}| = \sqrt{2} g^\prime \simeq 0.5$.
Therefore the right plot of Fig.~\ref{fig:susyLimits} resembles the second plot of Fig.~\ref{fig:MajoranaDM}. 
One can see that turning on $g_L$ makes the LHC constraint tighter.
The current HSCP LHC-8 TeV limit on the co-annihilation partner mass increases from 220 GeV to 300 GeV 
as $\theta$ changes from $\pi/2$ to $0$.
This is because the interaction strength of the $q \bar q \to (\gamma/Z)^* \to \phi^+ \phi^-$ process
increases due to inclusion of the $\rm SU(2)_L$ coupling found in $|D_\mu \Phi_L|^2$.   

\medskip

The dependences of the DM relic density and the lifetime of the co-annihilation partner on $\theta$ are more complicated,
and shown in Fig.~\ref{fig:theta}.
Here we plot $\Omega_{\rm DM} h^2$ (solid lines) and $\tau_\phi$ (dashed lines) as functions of $\theta$
by fixing $m_\chi = 300$ GeV
and varying $\Delta M = 1.2$, 1.4 and 1.6 GeV.
We see that $\Omega_{\rm DM} h^2$ is globally minimized at $\theta = 0$ and $\pi$ ($\phi = \phi_L$)
due to the relatively large $\rm SU(2)_L$ coupling. 
Another local 
minimum is found at $\theta = \pi/2$ ($\phi = \phi_R$).
The relic density has two local maxima
implying that there is a cancellation in $\langle \sigma_{\rm eff} v \rangle$ 
among $g_L$ and $g_R$ terms in Eq.~\eqref{eq:int_2}.   
The interference between $g_L$ and $g_R$ terms can also be observed
in the lifetime of $\phi$.
Unlike $\Omega_{\rm DM} h^2$, $\tau_\phi$ is minimized (maximized) at $\theta \simeq {{3 \pi} \over 8}$ 
(${{7 \pi} \over 8}$).

\subsection{Model 2: Scalar dark matter}
\label{sec:ScalarDM}

In this Section we consider Simplified Model~2 where the DM particle is a real singlet scalar, $\chi = S$, 
and the co-annihilation partner is a Dirac fermion, $(\eta^+, \eta^-) = (\overline{\Psi}, \Psi) = (\Psi^+, \Psi^-)$.
We take $\Psi$ to have the same quantum numbers as $\tau_R$ except for the $Z_2$ (dark sector) charge.
The Lagrangian is given as:
\beqn
\mathcal{L} & = & \mathcal{L_{\rm SM}} \,+\, \mathcal{L_{\rm DM}} \,+\, \mathcal{L_{\rm CAP}} \,+\, \mathcal{L_{\rm int}}, 
\nonumber \\
\mathcal{L_{\rm DM}} & = & \frac{1}{2} (\partial_\mu S)^2 - \frac{1}{2} m_{\rm DM}^2 S^2 \,,
\nonumber \\
\mathcal{L_{\rm CAP}} &=& \overline{\Psi} (i\slashed{D} - M_{\Psi}) \Psi \,,
\nonumber \\
\mathcal{L_{\rm int}} & = & 
g_{_{\rm DM}} \, S \, \overline{\Psi} \, P_R \, \tau \,+\, {\rm h.c.} \,,
\label{eq:Lag_2}
\eeqn
where $M_\Psi = m_{\rm DM} + \Delta M$ and $P_R = {{1 + \gamma_5} \over 2}$ is the right-handed projection operator
for Dirac spinors.
This simplified model can be realised 
for example in models with extra dimensions by regarding 
$\Psi$ 
as the first excited Kaluza-Klein (KK) mode of the $\tau$
and $S$ as a heavy and stable singlet, such as 
the first KK-mode of the Higgs boson \cite{Flacke:2008ne, Arrenberg:2008wy} or a scalar photon in $D \ge 6$ theories 
\cite{Arrenberg:2008wy, Dobrescu:2007ec}. 
In such models, the approximate mass-degeneracy, or a compressed spectrum between $m_{\DM}$ and 
$M_{\Psi}$, resulting in $\Delta M \ll m_{\rm DM}$, which is assumed in this paper, is 
justified because the mass of each of the KK modes for different 
particles is dominated by an universal contribution that is inversely proportional to the size of the extra dimension(s).
As in the case of Simplified Model~1a, this model is manifestly gauge invariant and renormalizable.

\medskip

We note that a term $|H|^2S^2$ is also allowed by the symmetry.
After the electroweak symmetry breaking, this term induces a 3-point interaction $hSS$
that gives the contribution to the direct detection
as well as $\Omega_{\rm DM} h^2$.
A phenomenological implication of this term has been 
well studied in the literature 
\cite{Djouadi:2011aa,LopezHonorez:2012kv,Djouadi:2012zc, Khoze:2014xha,Freitas:2015hsa,Brooke:2016vlw}.
Since the aim of this paper is to primarily study the effect of co-annihilation,
we simply assume that 
the coefficient of this term is small or otherwise exclude it from our simplified model.
\medskip

\begin{figure}[t!]
\begin{center}
\hspace{-5mm}
\includegraphics[scale=0.43]{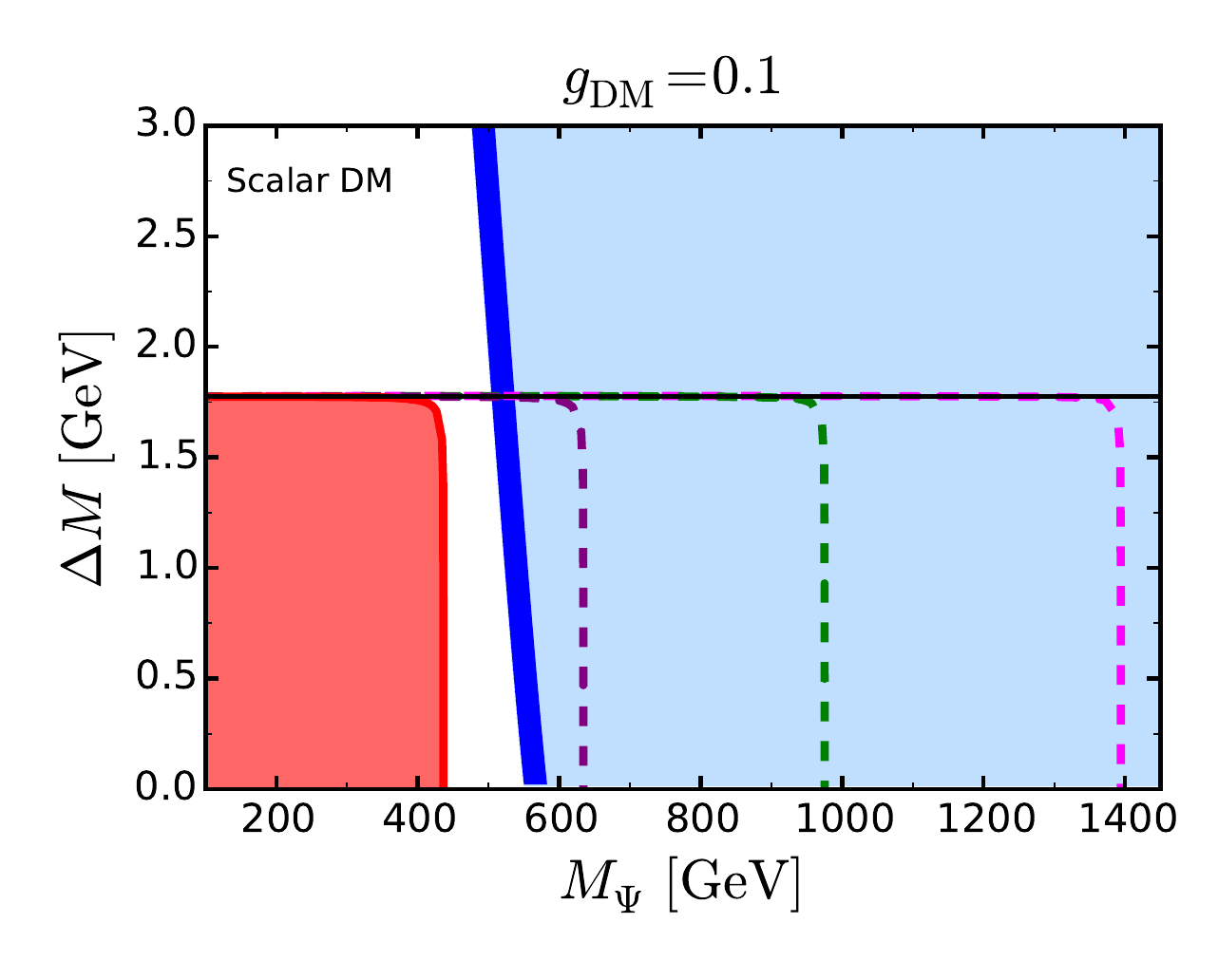}
\includegraphics[scale=0.43]{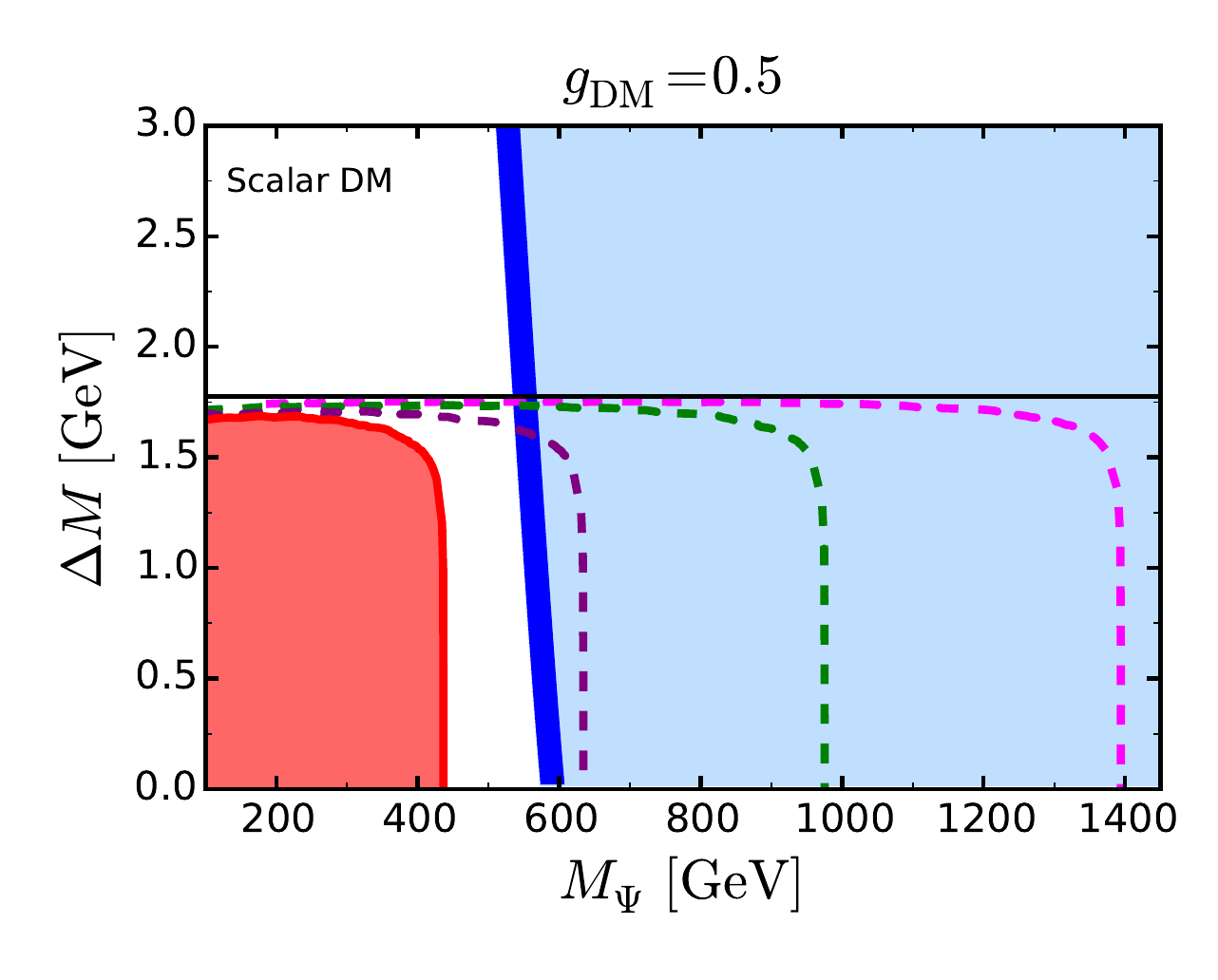}
\includegraphics[scale=0.43]{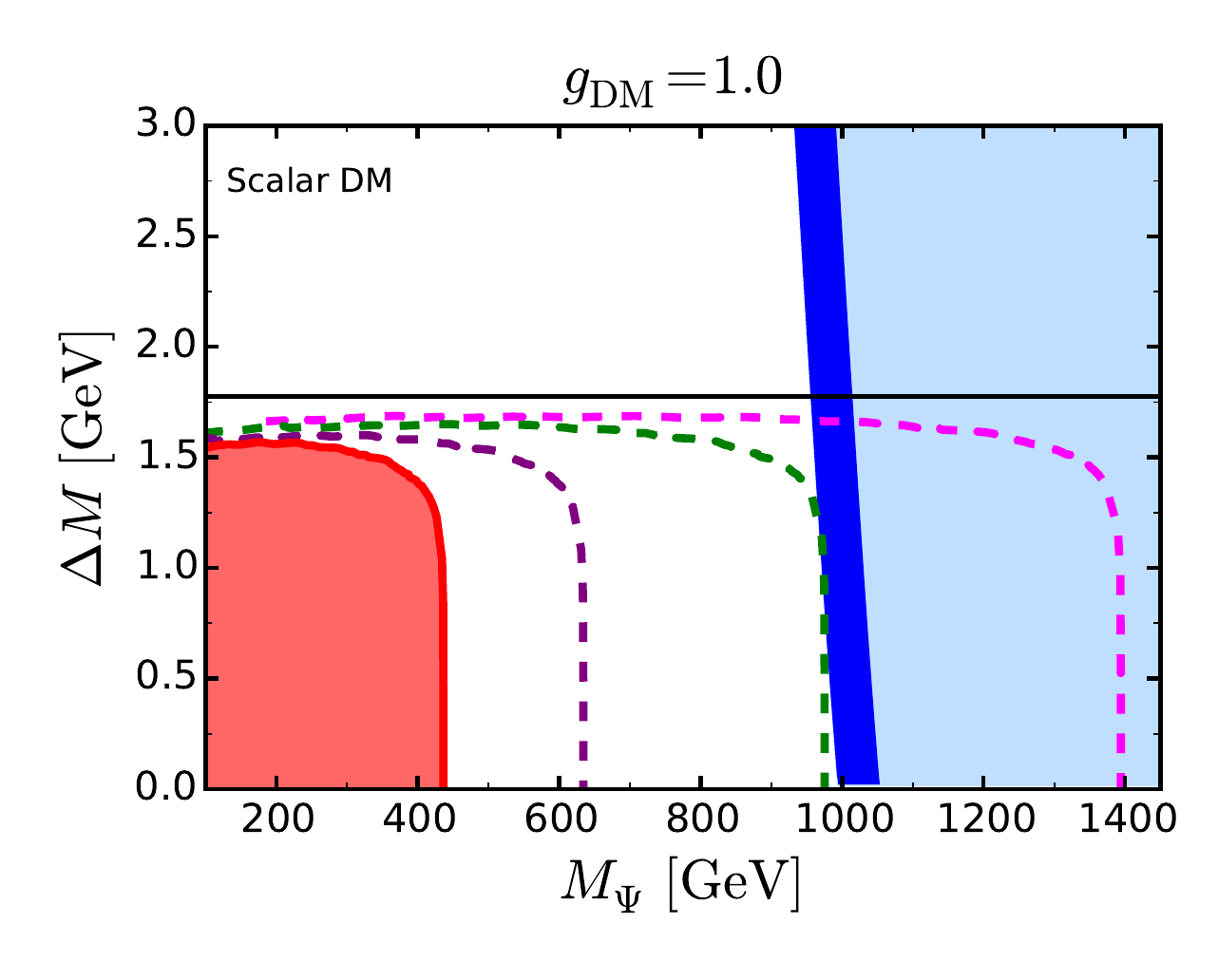}
\vspace{-3mm}
\caption{\label{fig:scalarDM} \small Model~2:
The co-annihilation strip and collider searches for scalar DM and a long-lived charged Dirac fermion $\Psi$. The dark-blue region satisfies the correct dark matter relic abundance within $3\sigma$, the light-blue region overproduces the dark matter energy density. The horizontal black line corresponds to the mass of the $\tau$ lepton. The region coloured in red corresponds to current HSCP limits for center-of-mass energy of 8 TeV and 18.8 fb$^{-1}$. The three dashed lines (purple, green and magenta) correspond to our projections for center-of-mass energy of 13 TeV and 30, 300 and 3000 fb$^{-1}$ of integrated luminosity respectively.}
\end{center}
\end{figure}

Fig.~\ref{fig:scalarDM} shows our numerical results of this simplified model for $g_{_{\rm DM}} = 0.1$, 0.5 and 1.0 from left to right.
Comparing it with Fig.~\ref{fig:MajoranaDM}, one can see that
the LHC limits are tightened but also the preferred co-annihilation partner mass by the relic density gets shifted to higher values.
This is because the number of degrees freedom for $\Psi$ is doubled compared to $\phi$.
Also, the production cross-section of the co-annihilation partners is enhanced compared to Model-1a because $q \bar q \to \Psi^+ \Psi^-$
does not incur velocity suppression near the threshold.
The current bound from the HSCP search excludes $M_{\Psi} \lesssim 410$ GeV 
and the projected sensitivity reaches 600, 950 and 1350 GeV for
the 13 TeV LHC with 30, 300 and 3000~fb$^{-1}$ integrated luminosity, respectively.
These current and projected limits are independent of $g_{_{\rm DM}}$ and $\Delta M$ as long as $\Delta M \lesssim 1.5$ GeV.

\medskip

\begin{figure}[t!]
\begin{center}
\includegraphics[scale=0.57]{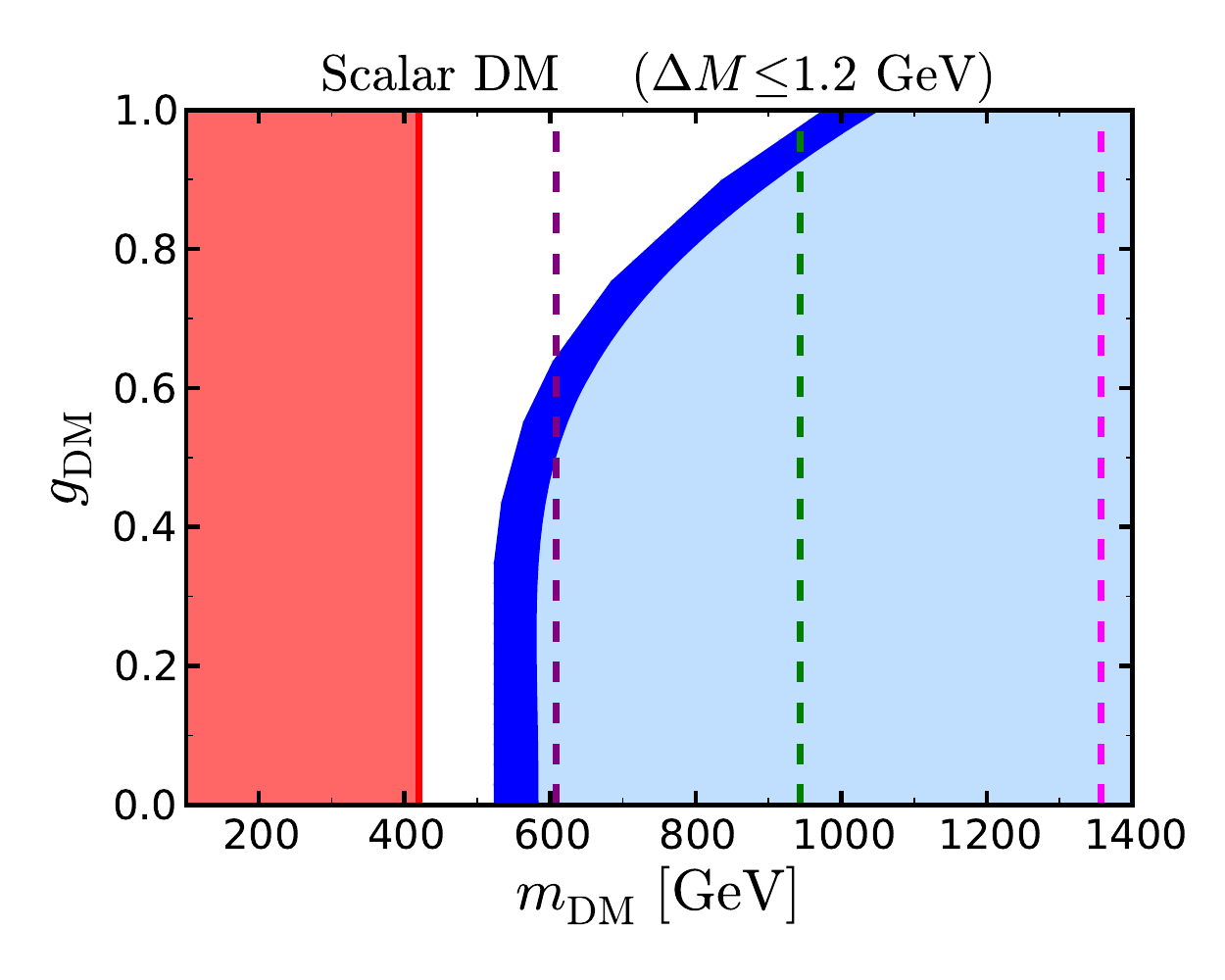}
\vspace{-5mm}
\caption{\label{fig:yRmDM} \small Model~2:  
Plot of the coupling $g_{_{\rm DM}}$versus the dark matter mass $m_{\rm DM}=m_{S}$. We scan over $\Delta M \in [0, 1.2{\rm \, GeV}]$, where $\Delta M\!=\!M_{\Psi}\!-\!m_{S}$. The dark blue band satisfies the correct DM relic abundance within $3\sigma$, the region in light blue overproduces the amount of DM. 
The colour-coding for the exclusion regions is the same as in the previous Figure.}
\end{center}
\end{figure}

The preferred co-annihilation partner mass required by the relic density (the dark-blue strip) is found around $M_\Psi \simeq$\,500$-$600 GeV
for $g_{_{\rm DM}} = 0.1$ and 0.5, and $M_\Psi \simeq$\,950$-$1050 GeV for $g_{_{\rm DM}} = 1.0$.
The impact of $g_{_{\rm DM}}$ and $m_{\rm DM}$ on $\Omega_{\rm DM} h^2$ can be seen more clearly in
Fig.~\ref{fig:yRmDM}, where limits from the LHC and $\Omega_{\rm DM} h^2$ are plotted 
in the ($m_{\rm DM}$, $g_{_{\rm DM}}$) plane scanning $\Delta M$ in the $[0, 1.2]$ GeV range.
In this plot, one can see the DM relic density is not sensitive to $g_{_{\rm DM}}$ until $g_{_{\rm DM}} \lesssim 0.5$.
This is because the $\langle \sigma_{\rm eff} v \rangle$ is determined by the process $\Psi^+ \Psi^- \to {SM\,{\rm particles}}$,
which is independent of $g_{_{\rm DM}}$.
For $g_{_{\rm DM}} > 0.5$,
the dependence enters through, i.e., $\Psi^\pm \chi \to {SM\,{\rm particles}}$ ($\langle \sigma_{\rm eff} v \rangle \propto g^2_{_{\rm DM}}$) and
$\Psi^\pm \Psi^\pm \to \tau^\pm \tau^\pm$ exchanging $S$ in the $t$-channel ($\langle \sigma_{\rm eff} v \rangle \propto g^4_{_{\rm DM}}$).
Considering the limit of the DM overproduction and the HSCP searches,
one can see that the entire parameter region with $g_{_{\rm DM}} \lesssim 1.0$ will be explored by
the LHC Run-2 with 3000 fb$^{-1}$ of integrated luminosity.

\subsection{Model 3: Vector dark matter}
\label{sec:VectorDM}

We now study the case in which the co-annihilation partner is a Dirac fermion, $(\eta^+, \eta^-) = (\overline \Psi, \Psi) = (\Psi^+, \Psi^-)$, as in Model-2 
but the dark matter is a neutral vector boson, $\chi = V_{\mu}$.
We modify the Lagrangian Eq.~\eqref{eq:Lag_2} with
\beqn
\label{eq:LagVector}
\mathcal{L}_{\rm DM} &=& \frac{1}{4} (\partial_\mu V_\nu - \partial_\nu V_\mu)^2 + \frac{1}{2} m_{\rm DM}^2 V_\mu V^\mu \,,
\nonumber \\
\mathcal{L}_{\rm int} &=& g_{_{\rm DM}} V^\mu \, \overline{\Psi}  \gamma_\mu  P_R \, \tau \,+\, {\rm h.c.}\,.
\eeqn
Similarly to Model-2, this simplified model can be realised in
models with extra dimensions by identifying $V_\mu$ as the KK photon and $\Psi$ as the KK $\tau$.
It may also be possible to interpret $V_\mu$ as a $\rho$ meson and $\Psi$ as a baryon in a new strong sector in composite models.  

\medskip

\begin{figure}[t!]
\begin{center}
\hspace{-5mm}
\includegraphics[scale=0.43]{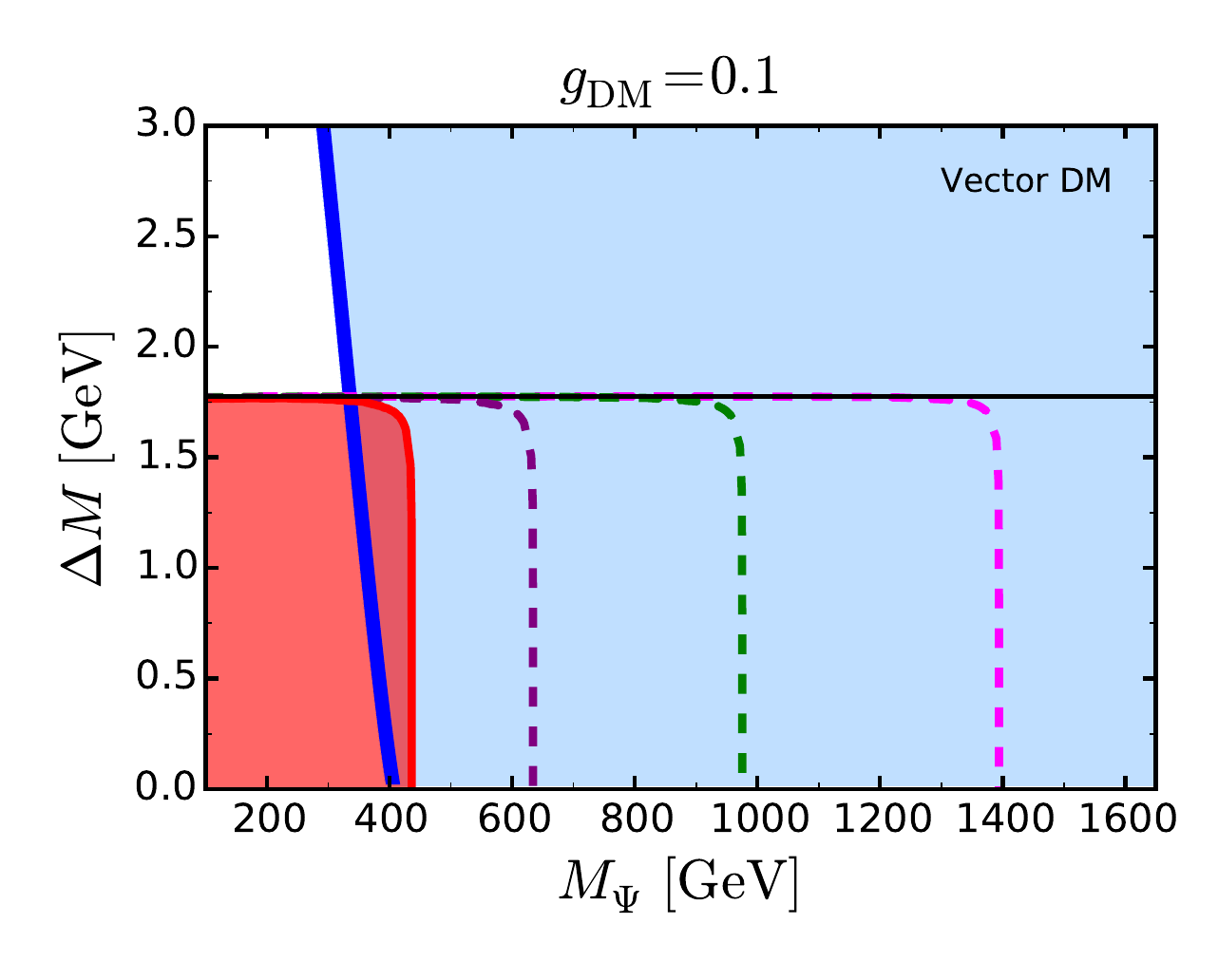}
\includegraphics[scale=0.43]{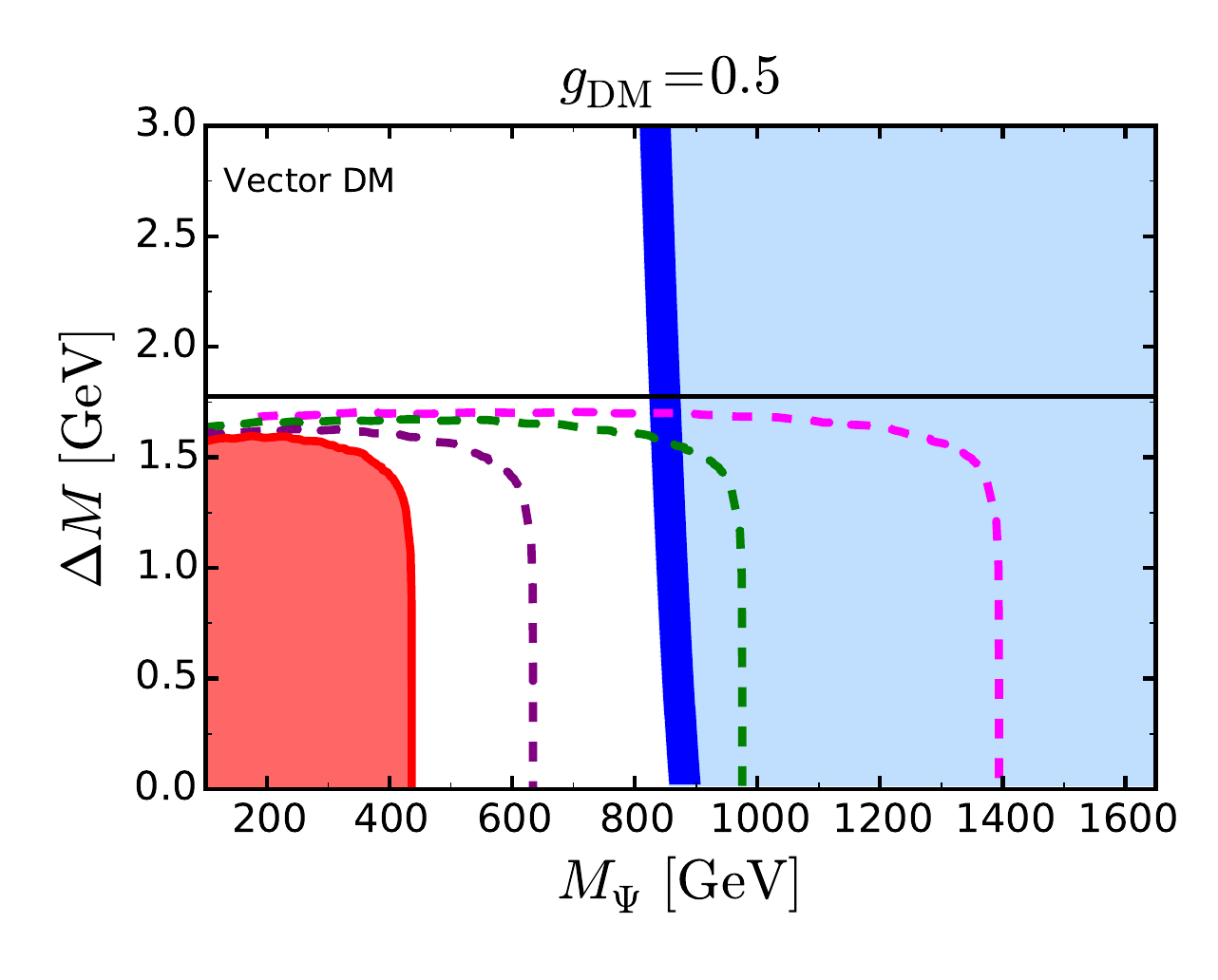}
\includegraphics[scale=0.43]{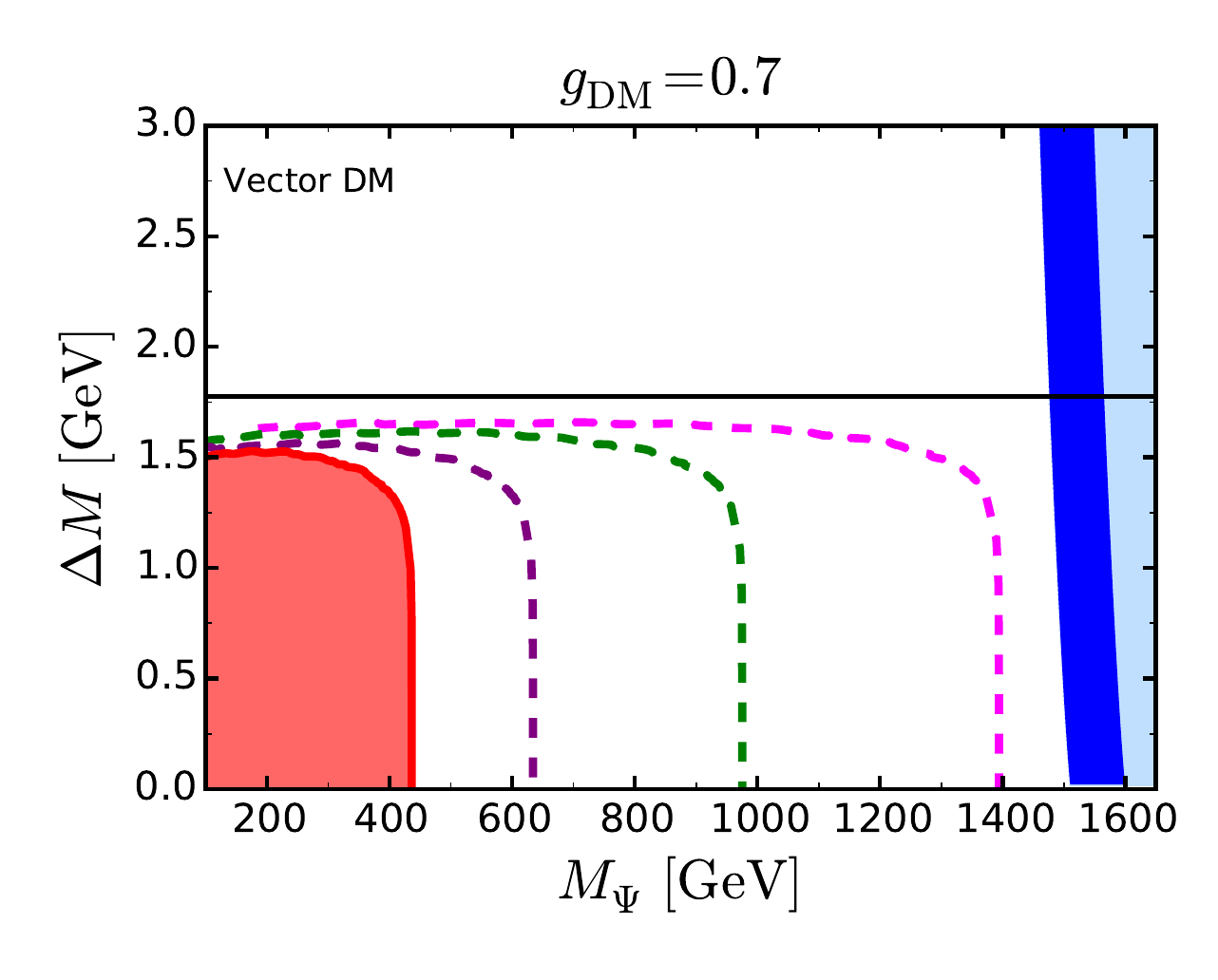}
\caption{\label{fig:vectorDM} \small Model~3: 
The co-annihilation strip and collider searches for vector DM and a long-lived charged Dirac fermion $\Psi$. The dark-blue region satisfies the correct dark matter relic abundance within $3\sigma$, the light-blue region overproduces the dark matter energy density. The horizontal black line corresponds to the mass of the $\tau$ lepton. The region coloured in red corresponds to current HSCP limits for center-of-mass energy of 8 TeV and 18.8 fb$^{-1}$. The three dashed lines (purple, green and magenta) correspond to our projections for center-of-mass energy of 13 TeV and 30, 300 and 3000 fb$^{-1}$ of integrated luminosity respectively.}
\end{center}
\end{figure}

\begin{figure}[t!]
\begin{center}
\includegraphics[scale=0.57]{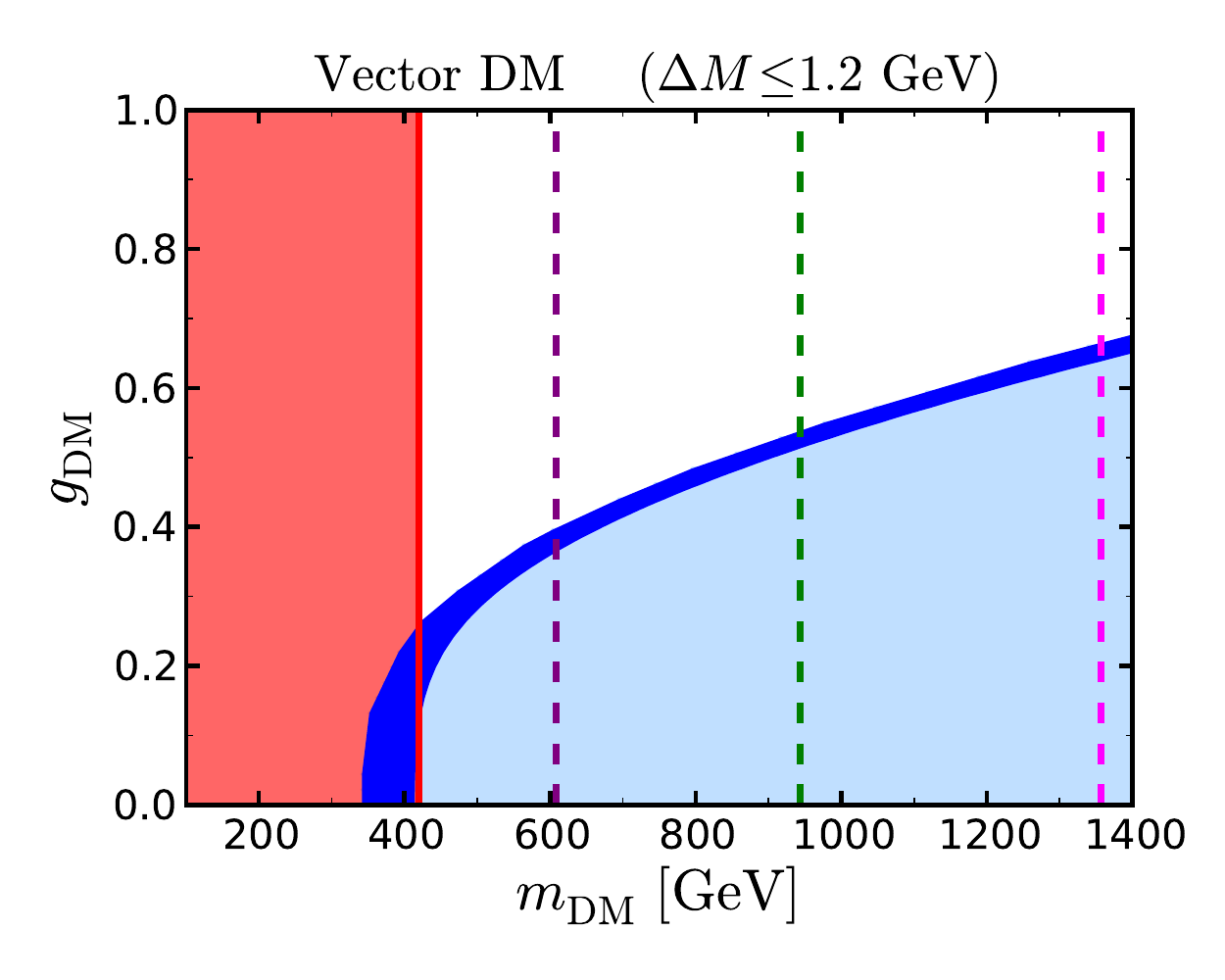}
\caption{\label{fig:gRmchi} \small Model~3: 
Plot of the coupling $g_{_{\rm DM}}$versus  the dark matter mass $m_{\rm DM}\!=\!m_{V}$. We scan over $\Delta M \in [0, 1.2{\rm \, GeV}]$, where $\Delta M\!=\!M_{\Psi}\!-\!m_{V}$, this is the mass region where the HSCP limits are independent of the coupling $g_{_{\rm DM}}$. The dark blue band satisfies the correct DM relic abundance within $3\sigma$, the region in light blue overproduces the amount of DM. The colour-coding for the exclusion regions is the same as in the previous Figure.}
\end{center}
\end{figure}

We show our numerical results of this model in Fig.~\ref{fig:vectorDM}, where $g_{_{\rm DM}} = 0.1$, 0.5 and 0.7
are examined from left to right. 
One can see that the current and projected LHC limits are almost identical to those found in Model-2,
since those models have the same co-annihilation partner $\Psi$,
and the relevant production process $q \bar q \to (\gamma/Z)^* \to \Psi \overline{\Psi}$ is independent of the spin of the DM.
On the other hand, the relic density constraint is quite different from the corresponding constraint in Model-2.
Interestingly, this model has larger $\Omega_{\rm DM} h^2$ for $g_{_{\rm DM}} = 0.1$ compared to Model-2.
In the limit $g_{_{\rm DM}} \ll 1$, Eq.~\eqref{eq:sigeff} implies 
\beqn
\frac{\langle \sigma_{\rm eff} v \rangle|_{\rm Model\,2}}{\langle \sigma_{\rm eff} v \rangle|_{\rm Model\,3}}
~ \simeq ~
\frac{({\tt g}_{_{V_\mu}} + {\tt g}_{_{\Psi}})^2}{({\tt g}_{_S} + {\tt g}_{_{\Psi}})^2} 
~ = ~
\frac{49}{25} \,.
\eeqn
On the other hand, for larger $g_{_{\rm DM}}$
the DM relic rapidly decreases, as can be seen in Fig.~\ref{fig:gRmchi}.
This is because the contribution of $V_\mu V_\mu \to \tau^+ \tau^-$
process is not chiral or velocity suppressed in this model
and it has a strong dependency on $g_{_{\rm DM}}$: $\langle \sigma(V_\mu V_\mu \to \tau^+ \tau^-) v \rangle \propto g^4_{_{\rm DM}}$.
One can see from Fig.~\ref{fig:gRmchi} that
a large region of the parameter space can be explored by the LHC and relic density constraints.
Nevertheless, the region with $m_{\rm DM} \gtrsim 1.4$ TeV and $g_{_{\rm DM}} \gtrsim 0.7$ 
may be left unconstrained even after the high luminosity LHC with 3000 fb$^{-1}$,
although such large values of $g_{_{\rm DM}}$ might
bring sensitivities for the direct and indirect detection experiments, which, however, is beyond the scope of this paper.

\section{Conclusions}
\label{sec:conclusions}

There is a considerable ongoing experimental and theoretical effort dedicated to the discovery of the dark matter. 
There has been a rapid development in the number and scope of direct and indirect detection experiments,
and in LHC and future collider searches of DM. A standard signature to search for dark matter at colliders is the mono-X 
(or multi-jets) plus missing energy. 
These searches are being exploited and interpreted in terms of simplified dark matter models with mediators. A growing number of the analyses are also dedicated to the direct search of the mediator
which can decay back to the SM degrees of freedom.

\medskip

In this article we considered an alternative DM scenario characterised by simplified models without mediators. 
Instead they include a co-annihilation partner particle in the dark sector. 
In the scenarios with a relatively compressed mass spectrum between the DM and its charged co-annihilation partner,
the latter plays an important role in lowering the dark matter relic density. 
The signal we study for collider searches is the pair-production of the co-annihilation partners 
that then ultimately decay into cosmologically stable dark matter. We have focused on the case when the dark matter candidate 
and the co-annihilation partner are nearly mass-degenerate, which makes the latter long-lived. Compared to other models of dark 
matter that rely on signals with missing energy at colliders, in these models the crucial collider signature to look for 
are tracks of long-lived electrically charged particles. 

\medskip

We have studied for the first time constraints from long-lived particles in the context of simplified dark matter models. 
We have considered three different scenarios for cosmological DM: a Majorana fermion, a real scalar and a vector dark matter. 
The model with Majorana DM can be motivated by theories with supersymmetry, such as the bino--stau co-annihilation strip in the MSSM. The model with vector DM can be motivated by Kaluza-Klein theories of extra dimensions, 
where the KK photon plays the role of dark matter. 
Nevertheless, in this work we have advocated for a simple (and arguably more inclusive) purely phenomenological approach 
and we have considered the couplings and the masses as free parameters.

\medskip

We have presented a set of simplified models which are complimentary to the standard mediator-based simplified DM models set,
and which can be used by the ATLAS and CMS experimental collaborations to interpret their searches for long-lived charged particles
to explore this new range of dark matter scenarios which we characterised in terms of 3 to 4 classes of simplified models 
with as little as 3 free parameters.

\section*{Acknowledgments}
This work is supported by STFC through the IPPP consolidated grant.  Research of VVK is supported in part by a Royal Society Wolfson Research Merit Award. ADP acknowledges financial support from CONACyT.
The work of KS is partially supported by the National Science Centre, Poland, under research grants
DEC-2014/15/B/ST2/02157 and DEC-2015/18/M/ST2/00054.


\appendix

\section{Indirect detection limits for Model~3}
\label{ap:2}

Unlike Model-1 and Model-2, Model-3 postulates a spin-1 dark matter particle, $V_\mu$.
The dark matter pair annihilation $V_{\mu} V_{\mu} \to \tau^+ \tau^-$ in the present universe is therefore not chiral suppressed and may be sensitive to indirect detection experiments. 
We compare the annihilation cross-section 
computed by \texttt{micrOMEGAs 4.1.5} with the upper limit derived 
from the gamma-ray observations of Milky Way dwarf spheroidal galaxies (dSphs)
at the Fermi-LAT satellite \cite{Ackermann:2015zua}.

\medskip

\begin{figure}[t!]
\begin{center}
\includegraphics[scale=0.7]{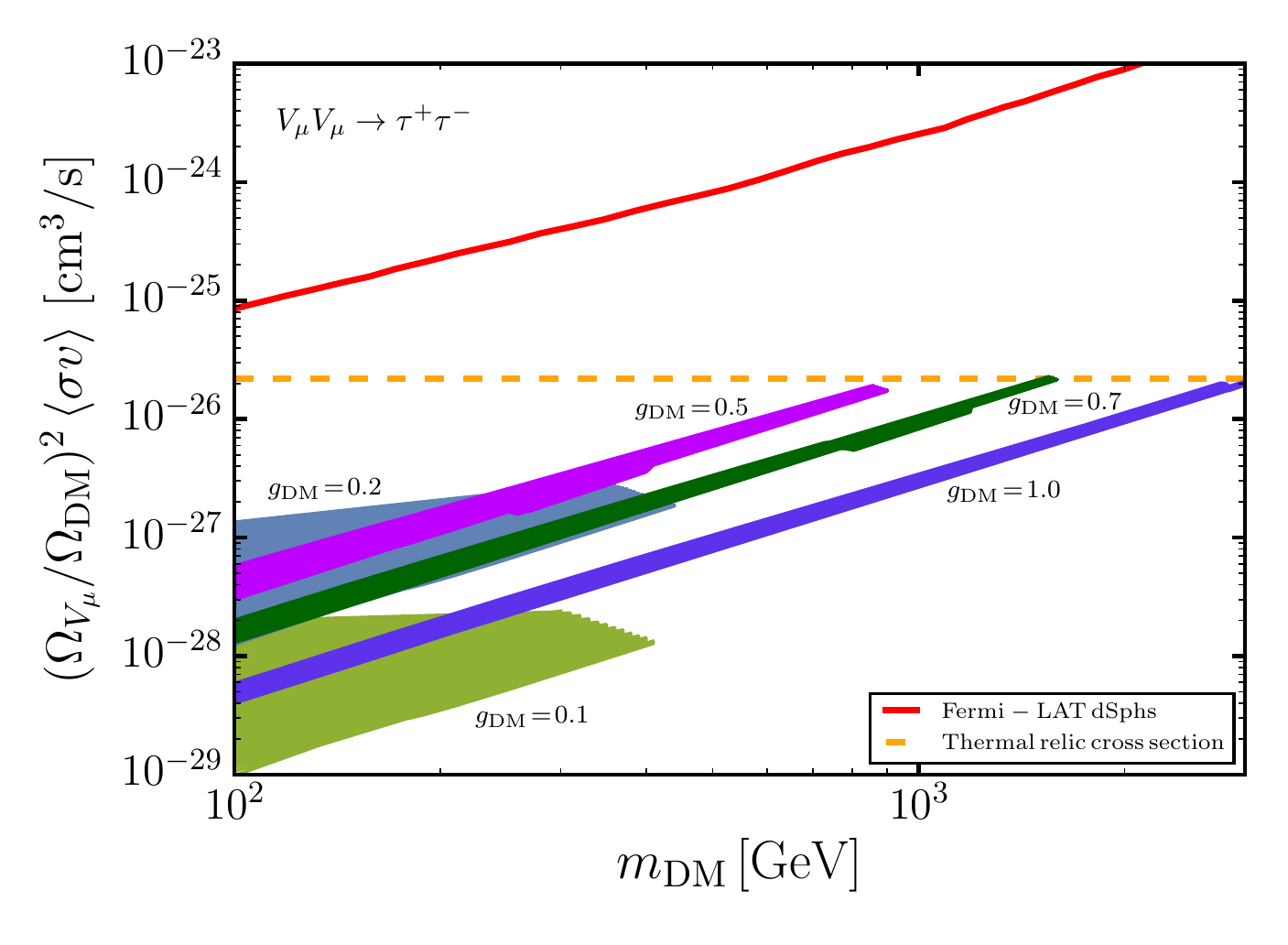} 
\vspace{-3mm}
\caption{\label{fig:indirect} \small
The rate of the dark matter annihilation $V_{\mu} V_{\mu} \to \tau^+ \tau^-$
as a function of the dark matter mass.
The red line corresponds to the current limit obtained by the gamma-ray observation of Milky Way dwarf spheroidal galaxies (dSphs) at the Fermi-LAT satellite \cite{Ackermann:2015zua}. 
The yellow dashed line corresponds to the thermal relic cross-section
assuming the pure $V_{\mu} V_{\mu} \to \tau^+ \tau^-$ process. 
The coloured regions correspond to different values of the coupling $g_{_{\rm DM}}$ and $\Delta M$ is scanned over the $[0, 3]$ GeV range.
}{}
\end{center}
\end{figure}

We show our results in Fig.~\ref{fig:indirect}, where  $\Delta M = M_{\Psi} - m_{\rm DM}$ is scanned over the $[0, 3]$ GeV range and  the coloured regions correspond to different values of the coupling $g_{_{\rm DM}}$, as explained in the figure.
In order to confront these with the experimental limit assuming the nominal DM flux,
these predictions are rescaled by the square ratio of the calculated relic abundance
and the observed one, $(\Omega_{V_\mu}/\Omega_{\rm DM})^2$ with $\Omega_{\rm DM} h^2 = 0.1197$.
We do not consider points that overproduce the relic abundance, i.e. all the points satisfy $\Omega_{V_\mu} h^2 \leq 0.1197$.

\medskip

As can be seen, by increasing the dark sector coupling $g_{_{\rm DM}}$ from 0.5 to 1.0,
the annihilation rate decreases.
This is because in this region, the abundance of $V_{\mu}$ is mainly determined by 
the same annihilation process $V_{\mu} V_{\mu} \to \tau^+ \tau^-$ in the early universe 
and $(\Omega_{V_\mu}/\Omega_{\rm DM})^2$ decreases more rapidly 
than the increase of the present time annihilation cross-section.
The situation is different for smaller values of $g_{_{\rm DM}}$,
where $\Omega_{V_\mu} h^2$ is determined by the co-annihilation mechanism
and the annihilation rate of 
$\,\Psi^+ \Psi^- \to {SM\,{\rm particles}}$, which does not depend on $g_{_{\rm DM}}$, as discussed in Section~\ref{sec:coan}.
One can therefore see that going from $g_{_{\rm DM}} = 0.1$ to $0.5$, the annihilation rate
increases. 

\medskip

The red line in Fig.~\ref{fig:indirect} shows the Fermi-LAT limit
assuming dark matter annihilation into the $\tau^+ \tau^-$ final state.
As can be seen, the predicted rate is more than two order of magnitude smaller than the current limit
across the parameter region.

\section{Limits in the mass vs lifetime plane}
\label{ap:1}

\begin{figure}[h!]
\centerline\\  \center
\scalebox{0.55}{\includegraphics{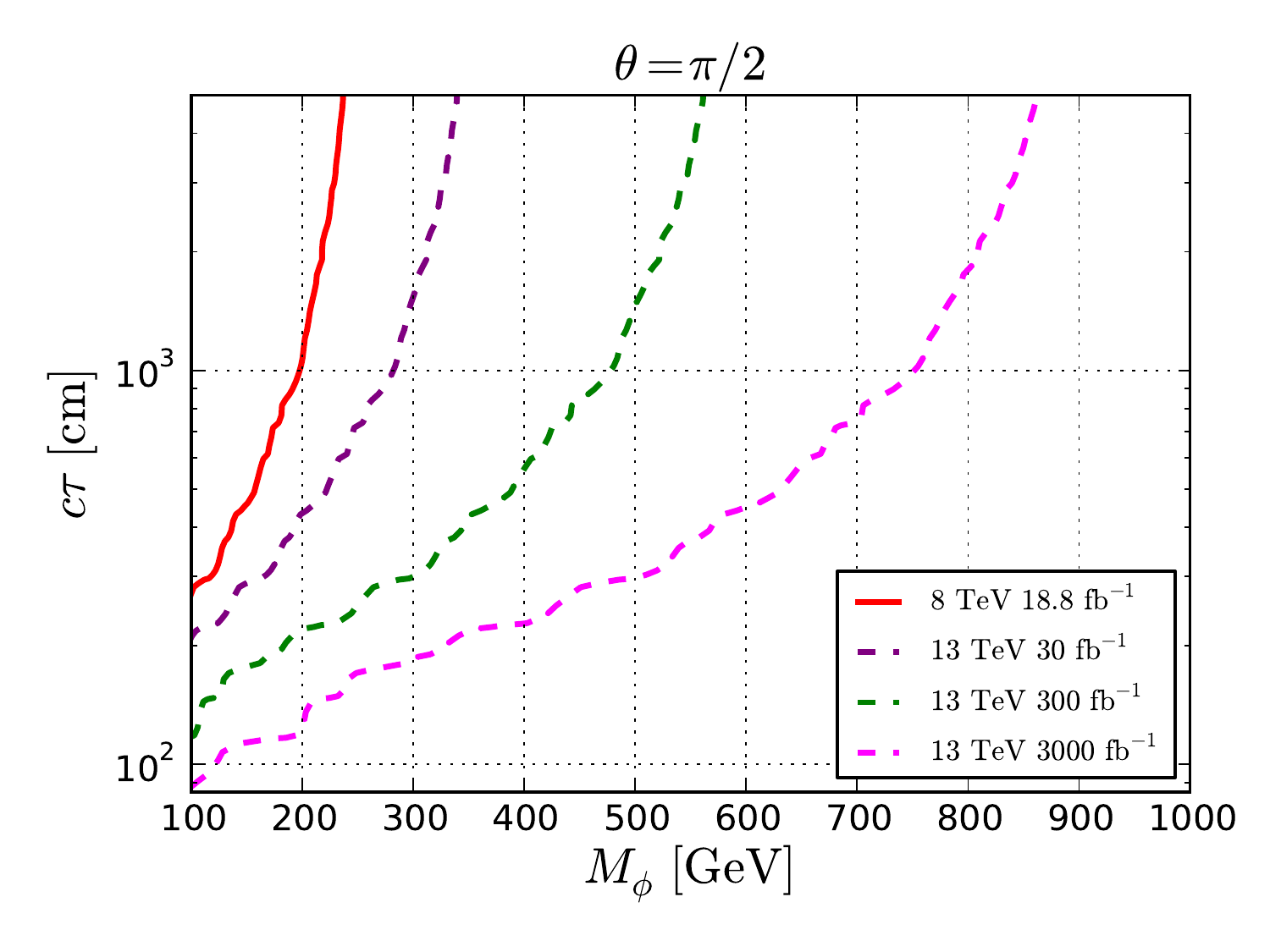}}
\hspace{-4mm}
\scalebox{0.55}{\includegraphics{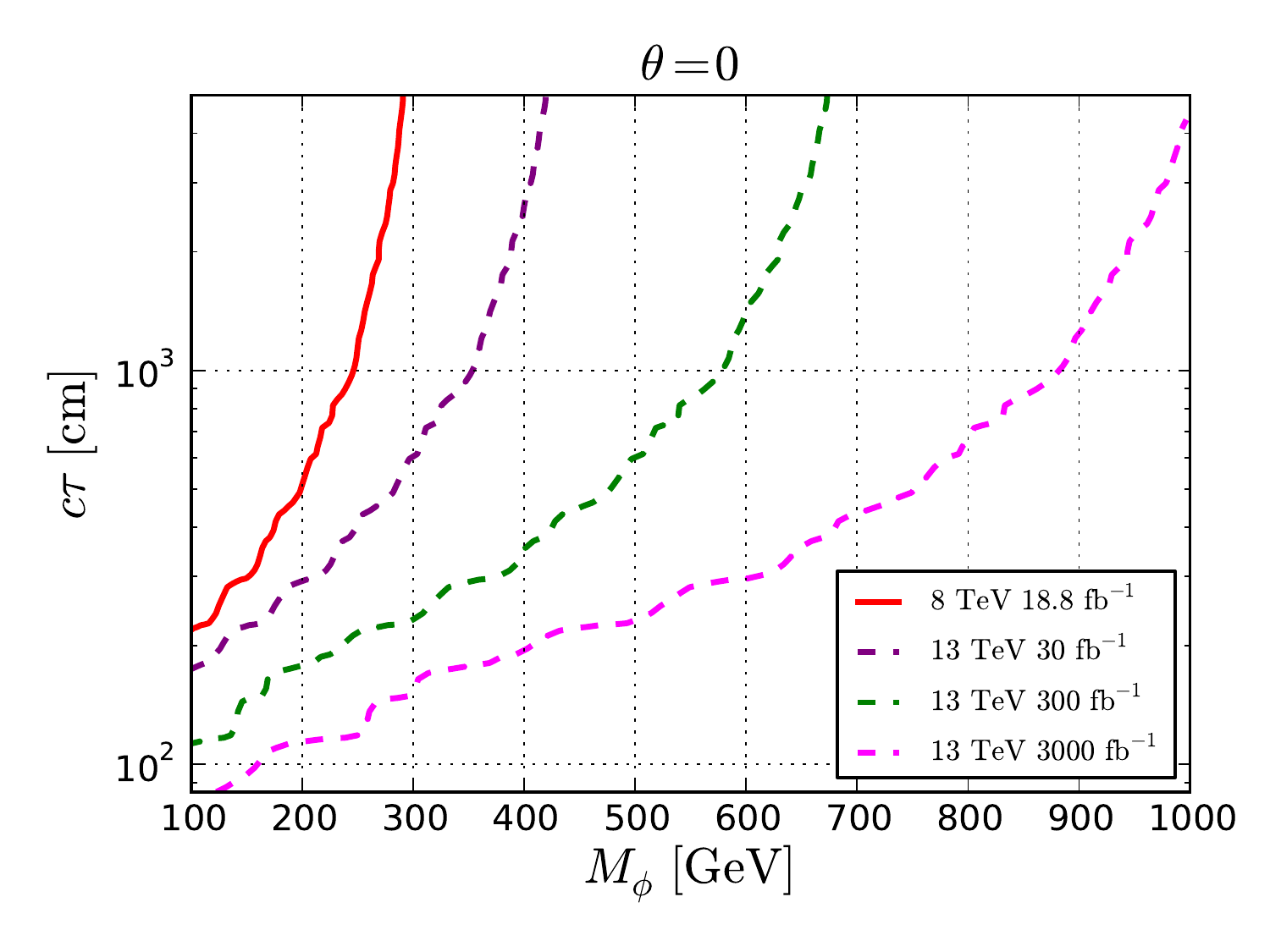}} 
\vspace{-3mm}
\caption{\label{fig:mlife1} \small
The 8 TeV (solid) and projected 13 TeV (dashed) limits from HSCP searches at the LHC
for pair-production of the scalar co-annihilation partner, $\phi^\pm$.
The projected limits correspond to the 13 TeV LHC with 30, 300 and 3000 fb$^{-1}$ integrated luminosities. 
}
\end{figure}

\begin{figure}[t!]
\begin{center}
\includegraphics[scale=0.5]{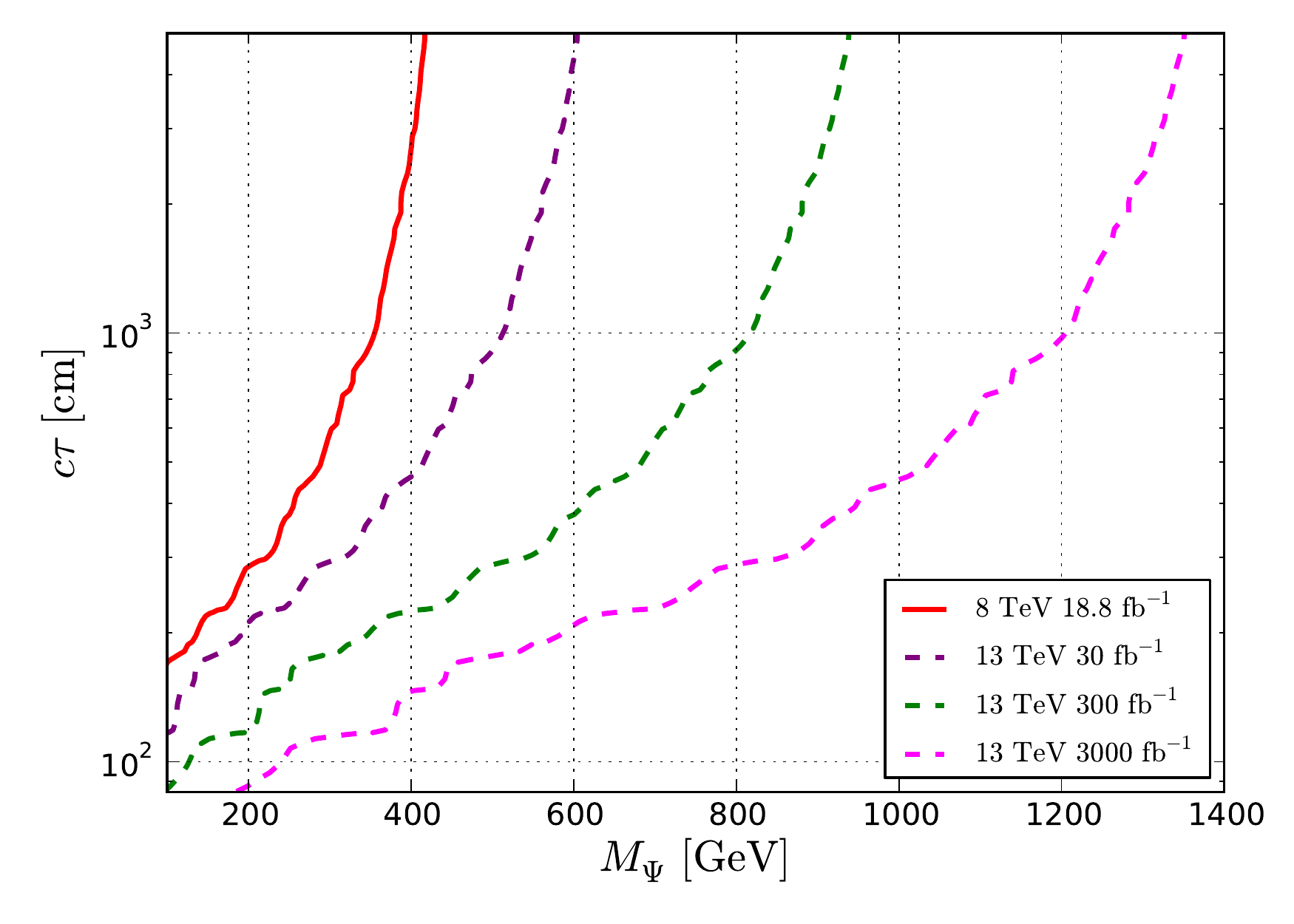} 
\vspace{-3mm}
\caption{\label{fig:mlife2} \small
The 8 TeV (solid) and projected 13 TeV (dashed) limits from HSCP searches at the LHC
for pair-production of the fermionic co-annihilation partner, $\Psi^\pm$.
The projected limits correspond to the 13 TeV LHC with 30, 300 and 3000 fb$^{-1}$ integrated luminosities. 
}
\end{center}
\end{figure}

The current and projected limits obtained from the heavy stable charged particle searches 
shown in Section~{\bf \ref{sec:res}} can also be presented in a more model-independent fashion  
by plotting on the mass vs lifetime plane.
The plots in Fig.~\ref{fig:mlife1} shows 
the 8 TeV (solid) and projected (dashed) limits 
for the pair-production of long-lived complex scalar field, $\phi$,
as a function of the mass, $M_\phi$, and the lifetime times the speed of light, $c \tau$.
The left plot assumes $\phi$ has the same quantum number as the right-handed $\tau$
corresponding to Simplified Model~1a.
In the right plot, on the other hand, the interaction of $\phi$ is obtained
by the procedure explained in Section {\bf \ref{sec:susy}} (Simplified Model~1b)
and taking $\theta = 0$.
The co-annihilation partner $\phi$ in this case corresponds to the purely left-handed stau
in SUSY theories.        
Fig.~\ref{fig:mlife2} shows the same limits for the fermionic co-annihilation partner, $\Psi$.
These limits are applicable for both Simplified Model 2 and 3 discussed in this paper.

\vspace{10mm}


\newpage


\footnotesize



\bibliography{SimplifiedDM}


\end{document}